\documentclass[osajnl,citeautoscript,twocolumn]{revtex4} 
\usepackage{amsfonts}
\usepackage{amsmath}
\usepackage{bm}
\usepackage{psfrag}
\newcommand{\R}{\text{I\!R}}
\newcommand{\ket}[1]{|#1\rangle}

\newcommand{\refcite}{~\onlinecite}
\newcommand{\ex}[1]{\mbox{e}^{#1}}
\newcommand{\With}{\quad\mbox{with}\quad}
\newcommand{\re}[1]{\text{Re}\left[#1\right]}
\newcommand{\im}[1]{\text{Im}\left[#1\right]}
\newcommand{\mf}{\mathfrak}
\newcommand{\vect}[2]{\begin{pmatrix}#1\\#2\end{pmatrix}}
\newcommand{\Vect}[3]{\begin{pmatrix}#1\\#2\\#3\end{pmatrix}}
\newcommand{\be}{\begin{equation}}
\newcommand{\ee}{\end{equation}}
\newcommand{\beq}{\begin{eqnarray}}
\newcommand{\eeq}{\end{eqnarray}}
\def\bal#1\eal{\begin{align}#1\end{align}}

\newcommand{\mat}[4]{\begin{pmatrix}
 #1& #2\\
 #3&#4
 \end{pmatrix}}

\begin{document}
\title{Polarization properties and dispersion relations for spiral 
resonances of a dielectric rod}
\author{Harald~G.~L.~Schwefel}
\author{A.~Douglas~Stone}
\email{Harald.Schwefel@yale.edu}
\affiliation{Yale University, Department of Applied Physics,  \\P.O.~Box 208284, New Haven, CT 06520-8284, USA} 
\author{Hakan~E.~Tureci}
\affiliation{Yale University, Department of Physics, \\P.O.~Box 
208120, New Haven, CT 06520-8120, USA}
%\date{\today}

\begin{abstract}
Dielectric microcavities based on cylindrical and deformed cylindrical shapes have been employed as resonators for microlasers. Such systems support spiral resonances with finite momentum along the cylinder axis. For such modes the boundary conditions do not separate and simple TM and TE polarization states do not exist. We formulate a theory for the dispersion relations and polarization properties of such resonances for an infinite dielectric rod of arbitrary cross-section and then solve for these quantities for the case of a circular cross-section (cylinder). Useful analytic formulas are obtained using the eikonal (Einstein-Brillouin-Keller) method which are shown to be excellent approximations to the exact results from the wave equation. The major finding is that the polarization of the radiation emitted into the far-field is linear up to a polarization critical angle (PCA) at which it changes to elliptical. The PCA always lies between the Brewster and total-internal-reflection angles for the dielectric, as is shown by an analysis based on the Jones matrices of the spiraling rays. 
\end{abstract}
%\ocis{060.2310  Fiber optics,
%260.5740  Resonance (Physical optics),
%260.5430  Polarization (Physical optics),
%080.2720  Geometrical optics, mathematical methods,
%140.4780  Optical resonators,
%230.3990  Microstructure devices,
%140.3410  Laser resonators,
%230.3990  Microstructure devices,
%260.2110  Electromagnetic theory,
%350.3950  Micro-optics,}

\ocis{060.2310,260.5740,260.5430,080.2720}
\maketitle

\section{Introduction}
There has been a great deal of recent interest in cylindrical and  deformed cylindrical dielectric resonators for micro-laser  applications~\cite{science98,nature97,rex02,Chern03,schwefel04}. From  the theory side there is a particular interest in the deformed case,  as in this case such resonators are wave-chaotic systems and can be  analyzed with methods from non-linear dynamics and semi-classical  quantum theory. Analysis of the resonances and emission patterns from  such systems has focused exclusively on the scalar Helmholtz equation  which describes the axial component of the electric (TM mode) or  magnetic (TE mode) fields for the case of resonant modes with zero  momentum in the axial direction ($z$-direction). For this case ($k_z  = 0$) the polarization state is unchanged by boundary scattering and  the non-trivial ray dynamics in the transverse plane does not affect  the polarization state of the resonant solutions. However it is  interesting to consider the solutions of the wave equation for both  cylindrical and deformed cylindrical dielectric rods with $k_z \neq  0$, since in this case the boundary scattering couples the electric  and magnetic fields and there no longer exist TM or TE solutions with  a fixed direction in space for one the fields. We refer to these  non-zero $k_z$ modes as ``spiral modes" and note that elastic  scattering from such spiral resonance modes has been measured previously by Poon et al.\ Ref.~\refcite{PoonCL98}, the polarization properties were unfortunately fully explored in those experiments.

The authors did however measure a systematic ``blue-shift" of the 
resonance modes with tilt angle, which has been predicted by Ref.~\refcite{Lock97b,RollS98} and which we will derive below.  The modes we 
study here are similar to the hybrid modes known in the study of 
optical fibers, where it is also well-known that there are no simple 
TE, TM or TEM-like modes, but more complex vector solutions are 
necessary.  Our emphasis differs however in several ways: 1) we are 
interested in uniform dielectric rods, not the variable index 
profiles typical of optical fibers.  2)  We are interested in modes 
which are not totally internally reflected so we can study the nature 
of the polarization of the emitted radiation in the far-field.  3) We 
are primarily interested in resonances of uniform rods with 
cross-sections in the range of tens to hundreds of $\mu$m, so they 
are strongly multi-mode and can be treated within the eikonal 
(semi-classical) approximation.
\begin{figure}[t]
\centering
\psfrag{local frame}{~~local coord.\ frame}
\psfrag{phi}{$\phi$}
\includegraphics[width=7cm]{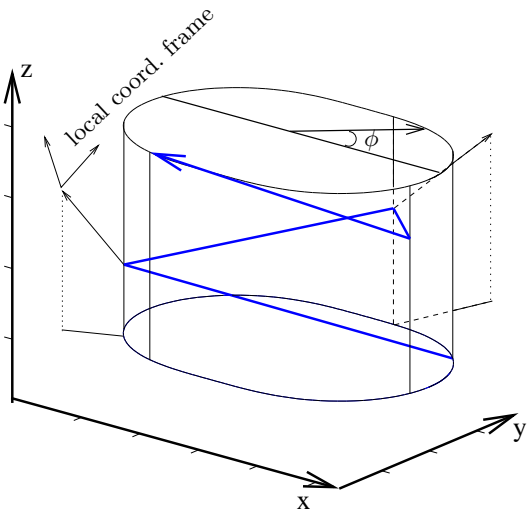}
\caption{Schematics of a spiraling ray in a rod with arbitrary cross-section. The ray can refract out and the polarization can be defined in the far-field with respect to a local coordinate system. For a general deformation, the polarization will change with the azimuthal angle $\phi$. Here, we will focus only on the circular cross-section, where the polarization is constant (with respect to 
$\phi$) when referred to the local coordinate system oriented along and perpendicular to a ray in the far-field.}
\label{fig:sprialQuad}
\end{figure}

Here, we will model finite resonators as infinite dielectric rods, 
neglecting the end effects in the $z$-direction; we therefore 
formulate the vector wave equation and the necessary boundary 
conditions for an infinite rod of arbitrary cross-section. Various 
approximations are possible to treat end effects when they are 
relevant, but we will not explore them here.  We define the resonant 
solutions (quasi-bound modes) of such a system and write down a 
general formalism which can be used to obtain exact numerical 
solutions for the vector resonances. We also show how the 
corresponding solutions can be used to derive the spatially-varying 
polarization state of the emitted radiation in the far-field.  We 
then study in detail these equations in the case of a circular 
cross-section (cylinder) for which a great deal of analytic progress 
and physical insight may be obtained. The analysis of the 
non-circular case for which wave-chaotic polarization states are 
possible will be published elsewhere~\cite{schwefel05b}.

\begin{figure*}[t]
\centering
\begin{minipage}[b]{8cm}
\psfrag{2eta}{$2\eta$}
\psfrag{a}{$\alpha$}
\psfrag{s}{$\sigma$}
\psfrag{chi}{$\chi$}
\psfrag{p1}{$\!\!\!\bm{p}^1$}
\psfrag{p2}{$\bm{p}^2$}
\psfrag{theta}{$\!\!\!\!\theta$}
A)\includegraphics[width=7cm]{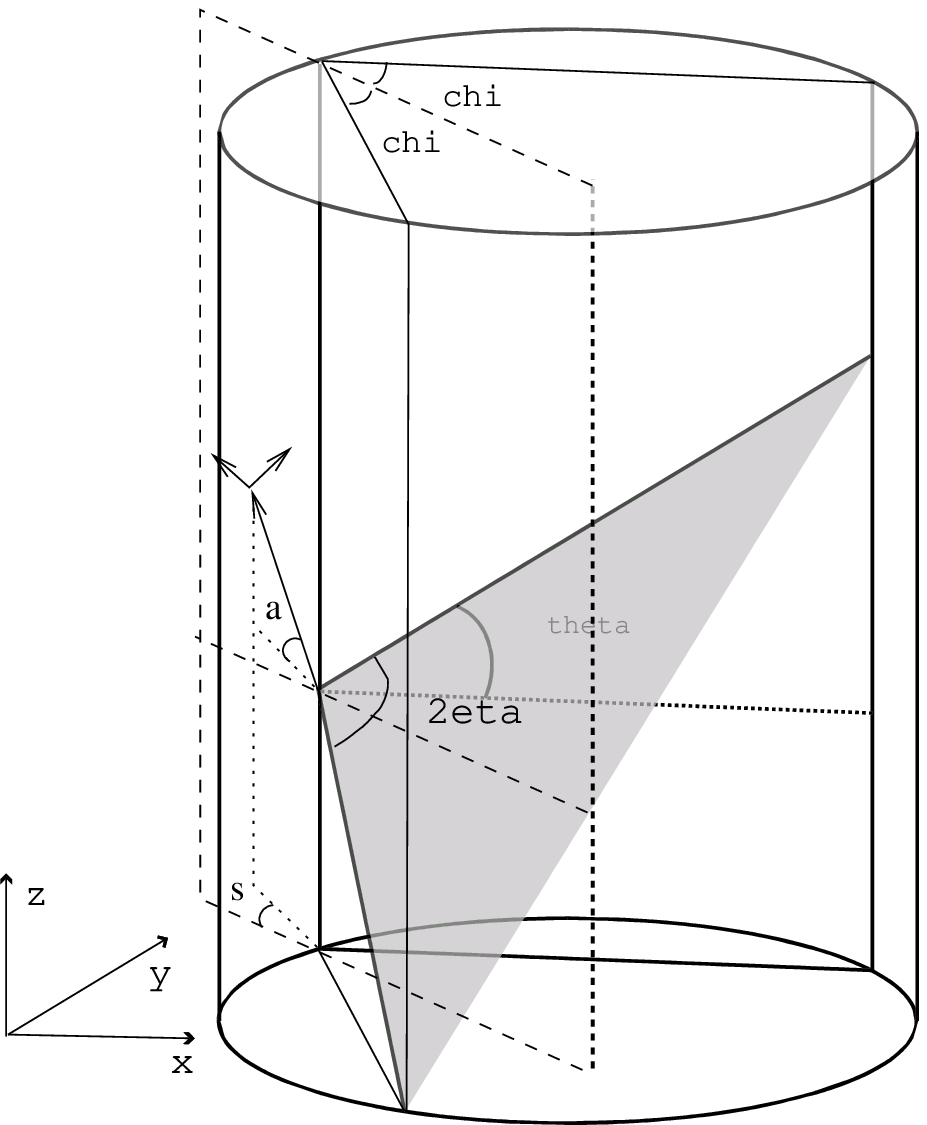}
%\caption{Schematic for the rays traveling in a 3-d cylinder.}\label{fig:3dschematic}
\end{minipage}\vspace{1cm }
\begin{minipage}[b]{7cm}
\psfrag{gamma}{\quad\quad\quad\quad$\gamma_1$}
\psfrag{kequal}{\quad\quad\quad$nk$}
\psfrag{kz}{$k_z$}
\psfrag{angle}{$\theta$}
B)\includegraphics[width=5cm]{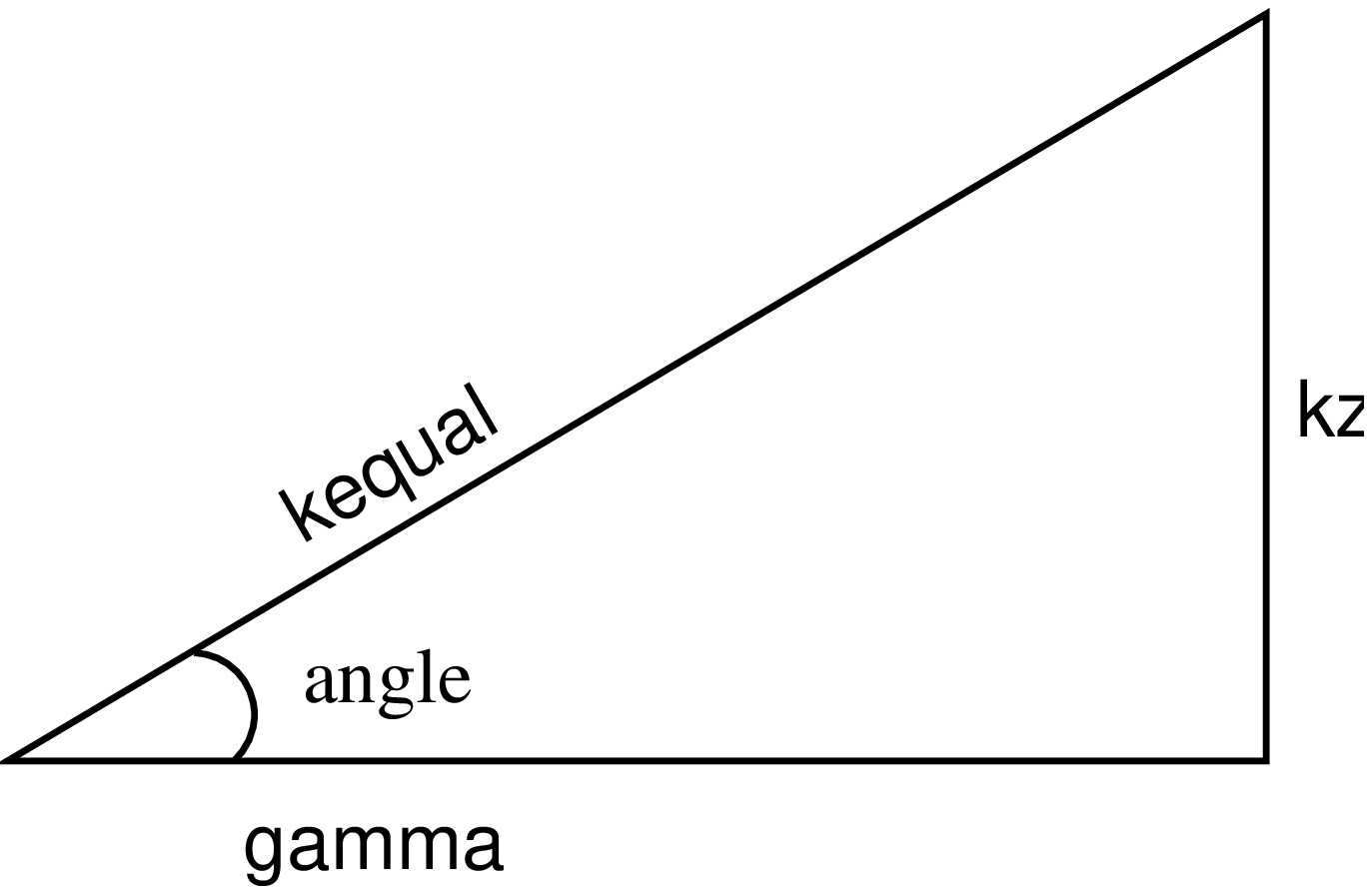}
\psfrag{gamma}{\quad\quad\quad\quad$\gamma_2$}
\psfrag{kequal}{\quad\quad\quad$k$}
\psfrag{kz}{$k_z$}
\psfrag{angle}{$\alpha$}
C)\includegraphics[width=5cm]{schemeKKz.eps}
%\caption{A), B) Schematic for the relations along the $z$-axis, inside (outside) of the cylinder of $k,k_z,\gamma_1,(\gamma_2)$ to the angles $\theta (\alpha)$. $\alpha$ is given by $n\sin\theta=\sin\alpha$ }
\end{minipage}
\caption{A) Coordinates used for ray dynamics in a rod with arbitrary  cross-section. The ray can refract out and the polarization can be  defined in the far-field with respect to a local cartesian coordinate  system tilted so that one of the axes is along the propagation  direction. We define $\eta$ as the angle of incidence in the plane of  incidence; $\chi$, the angle of incidence projected into the plane;  $\sigma$, the projected far-field angle; $\theta$, the tilt angle  measured from the cross-sectional $x-y$ plane, given by  $\tan\theta=k_z/\gamma_1$; $\alpha$, the external tilt angle  $\tan\alpha=k_z/\gamma_2 $; B), C) Schematics highlighting the  trigonometric relations among the inside and outside quantities. Note  that $\alpha$ can be found through the application of Snell's law,  $n\sin\theta=\sin\alpha$.}\label{fig:3dschematic}  
\end{figure*}

Our goal is to relate the polarization state in the far-field to the 
projected two-dimensional ray motion in the plane transverse to the 
$z$-axis. (See Fig.~\ref{fig:sprialQuad}) One can have spiral resonances 
which range from motion along the diameter of the rod in the 
transverse plane (bouncing ball type modes) to whispering gallery 
modes which circulate around the perimeter of the rod as they spiral 
along it. We will discuss how the polarization properties of the 
resonances vary as the angle of incidence in the plane ($\sin \chi$) 
and the spiral (tilt) angle ($\theta$) with respect to the $x-y$ 
plane varies. (In Fig.~\ref{fig:3dschematic} A) we introduce the 
relevant geometric parameters.) It should be noted that due to the 
curvature of the boundary, even modes which are totally-internally 
reflected according to geometric optics do emit by evanescent 
radiation into the far-field and their polarization fields can be 
obtained from exact solution of the wave equation, although 
experimentally it may be impractical to measure their weak emission 
far above the critical angle.  In the simplest case of $k_z=0$ 
$(\theta = 0)$ one finds pure linear polarization in the far-field; 
in addition, the resonant energies are just those of the 
two-dimensional problem of a dielectric disk.
In this two-dimensional case the resonant energies in the 
semiclassical limit are determined by the optical path length as well 
as by the phase shifts due to caustics and reflections at the 
boundary~\cite{tureci_thesis}. These boundary terms in the 
semiclassical limit correspond simply to the TIR phase shifts for TM and 
TE scattering off a plane dielectric interface when $\sin \chi > \sin 
\chi_c =1/n$ (here $n$ is the index of refraction of the rod 
surrounded by air); if  $\sin \chi < \sin \chi_c =1/n$ there is zero 
phase shift but just a loss (imaginary part of $k$) given by the 
Fresnel scattering coefficients. For the spiral modes ($k_z \neq 0$), 
the boundary terms have a new character corresponding neither to the 
TE nor to the TM Fresnel scattering and new phenomena can occur, such 
as a non-zero phase shift for modes which are not TIR. We derive 
below the generalization of the Fresnel scattering coefficients for 
spiral modes of the cylinder using the vector eikonal method. We find 
that the angle at which a non-zero phase shift sets in is always 
between the critical angle and the Brewster angle and coincides 
precisely with the onset of elliptical polarization in the far-field. 
We call this new key quantity the polarization critical angle (PCA)

In Section~\ref{sect:wave} of the paper we set up the relevant form 
of the vector wave equations for the infinite dielectric rod and 
formulate the boundary conditions for the quasi-bound modes 
(resonances).  In Section~\ref{sect:polarization} we discuss how to 
extract the far-field polarization of these quasi-bound modes.  All 
results are for the general case of arbitrary cross-section of the 
rod. In Section~\ref{sect:quasibound} we specialize to the case of 
the dielectric cylinder and reduce the resonance problem to a simple 
root-finding problem.  This equation is exact and is shown to yield 
the systematic blue-shift in Ref.~\refcite{PoonCL98}. In 
Section~\ref{sect:ebk} we reformulate the same problem using the 
eikonal (EBK) method which yields a simpler analytic formula for the 
resonances and allows a statement of the polarization problem in 
terms of generalized Fresnel coefficients. The exact and EBK 
resonance wavevectors are shown to agree quite well, down to small 
$k$.  In Section~\ref{sect:jones} we restate the polarization problem 
in terms of Jones matrices and thus derive the internal polarization 
state and the farfield polarization of the resonances in the 
semi-classical limit.  Both the Jones and EBK formulations are shown 
to yield the same answer for this quantity and to agree with the 
exact results to a good approximation. Finally, the origin of the 
polarization critical angle is explained.

\section{Wave equation and resonances for the infinite dielectric 
rod}\label{sect:wave}
For electromagnetic fields in free space interacting with uniform 
dielectrics, Maxwell equations reduce to the {\em vector Helmholtz 
equation}
\be
\left(\nabla^2 + n^2 k^2\right)
\left\{ \begin{matrix} \bm{E}(x,y,z)\\ \bm{B}(x,y,z) \end{matrix} \right\} =0,
\label{eq:vectorhheqn}
\ee
where $n$ is the uniform index of the rod and $n=1$ outside the rod. 
Thus $n$ differs from unity in an arbitrary closed simply-connected 
domain $\partial D$ in the $x-y$ plane for all values of $z$ (the rod 
is infinite in the $z$-direction).
The translational symmetry along the $z$-axis (see 
Fig.~\ref{fig:sprialQuad}) allows us to express the $z$-variation of 
the fields as
\be
\bm{E}(\bm{x}) = \bm{E}(x,y)\ex{-ik_z z},\quad
\bm{B}(\bm{x}) = \bm{B}(x,y)\ex{-ik_z z};
\ee
and henceforth the vectors $\bm{E},\bm{B}$ will refer to the $x,y$ 
dependent vector fields just defined. With this ansatz, we can show 
that the most general solution of the six-component vector wave 
equation for this problem is determined by $E_z$ and $B_z$ components 
alone; the perpendicular fields are given by linear combinations of 
these two scalar fields and their derivatives~\cite{schwefel_thesis}. 
Hence we must solve the two-component scalar wave equation
\be
\left(\nabla^2 + \gamma^2\right)
\left\{ \begin{matrix} E_z(x,y)\\ B_z(x,y)\end{matrix} \right\}
=0,\quad\mbox{with}\quad \gamma^2=n(\bm{x})^2 k^2-k_z^2,
\label{eq:EzBzHelmholz}\index{Helmholtz equation!vector reduced}
\ee
where we have introduced the {\em reduced wavevector} $\gamma$ which 
is the wavevector associated with the transverse  motion.
The complication of solving this remaining two-component Helmholtz 
equation stems from the fact that the two fields $E_z,B_z$ are 
coupled through the {\em boundary conditions}. Four independent 
boundary conditions are found through the application of the general 
Maxwell boundary conditions:
\beq
E_{z1}=E_{z2}\quad&\text{or}&\quad\partial_tE_{z1}=
\partial_tE_{z2}\label{eq:mBC1}\\
B_{z1}=B_{z2}\quad&\text{or}&\quad\partial_tB_{z1}=
\partial_tB_{z2}\label{eq:mBC2}\\
\frac{k}{\gamma_1^2}\partial_nB_{z1} - \frac{k}{\gamma_2^2}\partial_nB_{z2}
  &=&-\left(\frac{k_z}{\gamma_1^2}-\frac{k_z}{\gamma_2^2}\right)
  \partial_tE_{z1}\label{eq:mBC3}\\
\frac{n_1^2k}{\gamma_1^2}\partial_nE_{z1}-\frac{n_2^2k}{\gamma_2^2}\partial_nE_{z2}
&=&+\left(\frac{k_z}{\gamma_1^2}-\frac{k_z}{\gamma_2^2}\right)
\partial_tB_{z1}\label{eq:mBC4}.
\eeq
Here, $1$ and $2$ refer to inside and outside solutions. We recover 
the familiar special case of two-dimensional modes when we take 
$k_z=0$; in this case the boundary conditions can be satisfied with 
either $B_z=0$ (TM solutions) or $E_z=0$ (TE solutions) i.e. we get 
spatially uniform eigenpolarization directions. For the TM case we 
find  $\bm{E}_{\perp}=0$ and
\be
\bm{B}_{\perp}=\frac{i}{\gamma^2}n^2k\vect{-\partial_yE_z}{\partial_xE_z}
\ee
and for the TE case we find $\bm{B}_{\perp}=0$ and
\be
\bm{E}_{\perp}=-\frac{i}{\gamma^2}k\vect{-\partial_yB_z}
{\partial_xB_z}.
\ee
In both cases the equations Eqs.~\eqref{eq:mBC1}-\eqref{eq:mBC4} 
completely decouple and we have only to solve the scalar Helmholtz 
equation for $E_z$ or $B_z$ with the boundary conditions of the 
continuity of the field and its normal derivative on the boundary. 
In both cases the electromagnetic field is linearly polarized in the 
far-field, either in the direction parallel to the rod axis (TM) or 
perpendicular to it (TE).  We will now focus on the $k_z \neq 0$ 
modes which we will refer to as spiral modes. We will not consider 
the extreme case where $k=k_z$ (TEM modes).

Our interest is in the dielectric rod as a resonator, i.e. as a 
device for trapping light. Experiments on resonators fall into two 
broad categories, and the presence of quasi-bound modes are 
manifested differently in these two situations. One can measure 
elastic scattering of incident laser light from such a rod, as was 
done in Ref.~\refcite{PoonCL98} and focus on the specific wavevectors 
at which one observes scattering resonances.  For this case the 
linear wave equation that we are studying provides an exact 
description.  One can also imagine the rod containing a gain medium 
and when pumped emitting laser light into these spiral 
resonances~\cite{tureci_thesis}. In such a case, the linear wave 
equation is not an exact description; however for high-$Q$ resonances 
typically the resonances of the passive and active cavity are very 
similar~\cite{HarayamaDI03}.  In order for the laser to emit specifically 
into spiral modes, there would have to be some mechanism to suppress 
lasing of planar ($k_z=0$) modes; one could imagine doing this with a 
small seed pump which is tuned to the frequency of a spiral mode and 
pushes it above threshold before all the other modes. It is possible 
to describe these two different physical situations (elastic 
scattering and lasing) using appropriate boundary conditions on the 
linear vector Helmholtz equation. To be precise, we assume that the 
rod is bounded by the interface $\partial D$ given by (see 
Fig.~\ref{fig:sprialQuad})
\be
\partial D=R(z,\phi)\quad\forall\quad z\in\R,\quad\phi\in[0,2\pi].
\ee
We are typically interested in boundaries which are smooth and not 
too far from a circle, hence it is natural to expand the internal and 
external solutions $E_z$ and $B_z$ of Eq.~\eqref{eq:EzBzHelmholz} in 
cylindrical harmonics~\cite{Tureci05}
\bal
\vect{E_z^<}{B_z^<}&=\sum_{m=-\infty}^\infty
\left[\vect{\alpha_m}{\xi_m}H^+_m(\gamma_1r)+\vect{\beta_m}{\eta_m}H^-_m(\gamma_1r)\right]
\,e^{im\phi}\label{eq:GeneralAnsatz}\\
\vect{E_z^>}{B_z^>}&=\sum_{m=-\infty}^\infty
\left[\vect{\upsilon_m}{\zeta_m}H^+_m(\gamma_2r)+\vect{\delta_m}{\vartheta_m}H^-_m(\gamma_2r)\right]
\,e^{im\phi},\label{eq:GeneralAnsatz2}
\eal
where $H^\pm$ are the Hankel-functions, $H^-$ representing an 
incoming wave from infinity and $H^+$ an outgoing wave.

\begin{figure}[t]
\psfrag{iphi}{$I(170^{\circ})$}
\psfrag{rekR}{$\re{kR}$}
\psfrag{imkR}{$\im{kR}$}
\centering
\includegraphics[width=7cm,height=4cm]{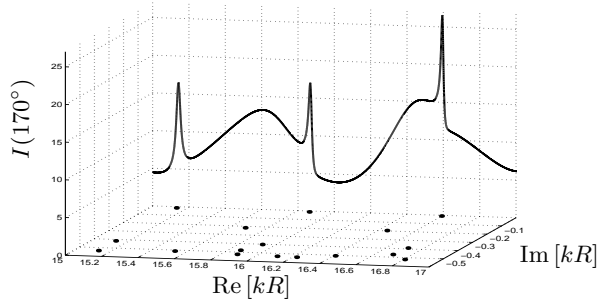}
\caption{A comparison of scattering and emission pictures for 
quasi-bound modes. The complex quasi-bound mode frequencies are 
plotted on the $\re{kR}-\im{kR}$ plane. On the back panel we plot the 
real $k$ $S$-matrix, scattering cross-section at $170^\circ$ with 
respect to the incoming wave direction. Notice that the most 
prominent peaks in scattering intensity are found at the values of 
$k$ where a quasi-bound mode frequency is closest to the real-axis. 
These are the long-lived resonances of the cavity. Also visible is 
the contribution of resonances with shorter lifetimes (higher values 
of $\im{kR}$) to broader peaks and the scattering background. 
Calculations are for a dielectric cylinder with $n=1.5$ and for 
$k_z=0$.}
\label{fig:scattering}
\end{figure}

To describe the scattering experiment we simply apply the boundary 
conditions Eqs.~\eqref{eq:mBC1}-\eqref{eq:mBC4} which will connect 
these interior and exterior solutions by a set of linear equations 
for the coefficients $\alpha_m, \xi_m, \ldots, \vartheta_m$.  These 
linear equations will have solutions for any incident wavevector $k$ 
and will define a scattering matrix for the fields which can be used 
to calculate the intensity scattered at any given far-field angle. 
An example of such a calculation for the cylinder is given in 
Fig.~\ref{fig:scattering}; the values of $kR$ ($R$ is the cylinder 
radius) at which rapid variation is observed are the resonant 
wavevectors for which the incident light is trapped for long periods. 
However the precise pattern of radiation in the far-field in this 
case is determined by both the scattered and incident radiation and 
is not representative of a lasing mode for which there is no incident 
radiation.  To determine the resonances corresponding to emission 
from a source it is conventional to use the Sommerfeld or radiation 
boundary conditions; in this case we would set the incident waves 
from outside to zero (the coefficients $\delta_m,\vartheta_m$ in 
Eq.~\eqref{eq:GeneralAnsatz2}) and still impose the boundary 
conditions of Eqs.~\eqref{eq:mBC1}-\eqref{eq:mBC4}.  The resulting 
linear equations for the remaining coefficients would not have any 
solutions for real $k$ (since current is not conserved), but would 
have discrete solutions for complex $k$ values; those solutions are 
known as the quasi-bound modes of the problem.  It can be shown that 
the discrete complex solutions of this problem correspond to the 
poles of the $S$-matrix of the current-conserving problem when that 
$S$-matrix is continued to complex values of $k$ (see e.g.\ 
Ref.~\refcite{tureci_thesis,Tureci05}). $\re{k}$ at the pole position 
gives the approximate location of the resonant peak in the on-shell 
$S$-matrix and the $\im{k}$ gives the width of the peak; the 
$Q$-value of the resonance is just $Q=2 \re{k}/|\im{k}|$.  These 
statements hold for $k_z =0$ resonances; for the spiral modes there 
are solutions for all $k_z$ and for discrete complex values of the 
reduced wavevector $\gamma$.  Once the resonant wavevector has been 
found the coefficients of the outgoing waves can be determined to 
give the emission pattern and polarization properties of the emitted 
radiation.  As is well-known, the Hankel functions with complex $k$ 
and large argument grow exponentially and do not provide normalizable 
fields at infinity.  This is unimportant for studying the emission 
patterns as a function of far-field angle or polarization properties, 
although it may cause some practical difficulties in numerical 
algorithms.  If desired, this unphysical feature of the solutions can 
be avoided for a given resonance by adding a tunable imaginary part 
of the index of refraction to yield a solution with real $k$ outside 
the dielectric.  This imaginary part represents linear amplification 
in the medium and would give an estimate for the lasing threshold for 
that mode if mode competition effects were negligible.  One finds 
that these real $k$ solutions are continuously related to the complex 
$k$ quasi-bound states at real index and have approximately the same 
spatial properties except for the absence of growth at infinity.

\section{Polarization of spiral modes in the 
far-field}\label{sect:polarization}
Having outlined how to solve for the quasi-bound spiral modes we now 
analyze their polarization properties. Strictly speaking polarization 
of the time-harmonic electromagnetic field cannot be defined inside 
the cavity or in the near field around it as the electric and 
magnetic fields need not be perpendicular to one another or to the 
direction of energy flow.  In the far-field on the other hand, where 
the radiation is well-approximated locally by a plane wave, we should 
be able to analyze the polarization of the emission from spiral modes 
in conventional terms.   Assuming the matching problem is solved, the 
coefficients $\upsilon_m,\zeta_m$ determining $E_z^<(\rho,\varphi), 
B_z^<(\rho,\varphi)$ are known and these fields can be differentiated 
in order to find $\bm{E}_\perp$ everywhere outside the rod.  These 
relations simplify if we use the large argument expansion of the 
Hankel functions and their recursion relations to find the relative 
magnitudes of the $\bm E$-field components in cylindrical coordinates 
as $\rho \rightarrow \infty$,
\be
%|\bm{E}|^2=
\left|\!\!\Vect{E_\rho}{E_\varphi}{E_z}\!\!\right|^2\!\!\!\sim
\left|\!\!\Vect{k_z/\gamma_2\sum\upsilon_me^{ im(\varphi-\pi/2)}}
{\hfill k/\gamma_2\sum\zeta_me^{ im(\varphi-\pi/2)}}
{\hfill\sum\upsilon_me^{ im(\varphi-\pi/2)}}\!\!\right|^2\!\!\!
=\left|\!\!\Vect
{ k_z/\gamma_2E_z}
{\hfill k/\gamma_2B_z}
{\hfill E_z}\!\!\right|^2
\ee
To extract the polarization at a particular angular direction 
$\varphi$ we need to recognize that far away from the rod the 
radiation is not propagating in the radial ($\rho $) direction with 
respect to the cylindrical coordinates centered on the rod axis but 
is instead propagating at angle $\alpha$ between the $\rho$ and $z$ 
directions determined by Snell's law for the $z$ motion (See 
Fig.~\ref{fig:3dschematic}). We can then rotate our coordinate system 
by $\alpha$
\be
\begin{pmatrix}
\cos\alpha&0&-\sin\alpha\\
0&1&0\\
\sin\alpha&0&\cos\alpha\end{pmatrix}
\begin{pmatrix}
\tan\alpha E_z\\
\sec\alpha B_z\\
E_z\end{pmatrix}
=\Vect{0}{\sec\alpha B_z}{\sec\alpha E_z}.
\ee
In this rotated coordinate system in the farfield, the electric field 
(on the RHS) only has two components in the plane transverse to the 
propagation direction. Thus, the polarization of the electric field 
in the far-field is then determined by the ratio of these two 
components, $E_z(\varphi) /B_z (\varphi)$.  If these two field 
amplitudes have zero phase difference we have linear polarization in 
a certain direction which can vary as the angle of observation 
$\varphi$ varies; if there is a non-zero phase shift  $\Delta$ 
between them, then we typically have elliptical polarization except 
in the specific case of $\Delta=\pi/2$ and $|B_z|^2=|E_z|^2$ 
corresponding to circular polarization.  It should be noted that when 
the angle $\alpha$ is complex $k_z > k$ then the wavevector 
$\gamma_2$ of the outgoing Hankel functions is pure imaginary and 
there is no propagating radiation as $\rho \rightarrow \infty$; this 
corresponds to the total internal reflection condition with respect 
to the $z$ motion of internal ray in the cylinder, $n\sin \theta = 1$ 
(see Fig.~\ref{fig:3dschematic}), for which there is no evanescent 
escape.  It should be emphasized however that this does not 
correspond to the true total reflection condition for spiraling rays, 
which comes at smaller $\theta$ except in the case of normal 
incidence in the transverse plane.  In particular one can ask about 
the polarization states in the far-field of spiral whispering gallery 
modes, which emit solely by evanescent escape.

The analysis up to this point has been exact for a dielectric rod of 
arbitrary cross-section and the various formulas can be used to solve 
for both the resonance wavevectors and polarization properties of the 
spiral modes of such a system. We have developed and implemented a 
numerical algorithm to do this~\cite{schwefel_thesis}; we intend to 
describe the algorithm and present results for deformed cylinders in 
a subsequent paper~\cite{schwefel05b}.  At this point we specialize 
to the problem of spiral modes of a cylinder for which a number of 
analytic techniques are possible which will allow us to develop a 
useful physical picture.

\section{Quasi-bound resonances in the cylinder}\label{sect:quasibound}
\begin{figure}
\centering
\psfrag{m}{$m$}
\psfrag{reg}{\hspace{-2ex }\raise1ex\hbox{$\re{\gamma_1}$}}
\includegraphics[width=7cm]{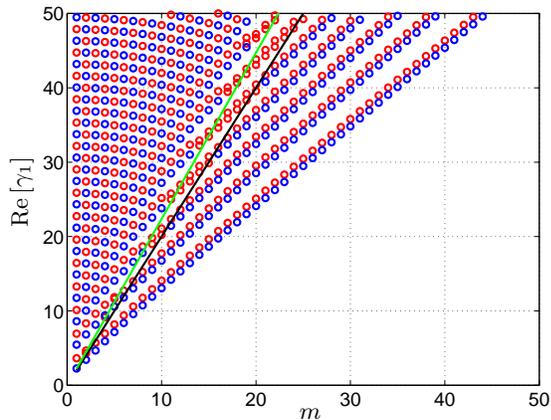}
\caption{TE like resonances (red circles) and TM like resonances 
(blue circles) for a cylinder with $n=2, \theta=0.2$, dashed 
line corresponds to the critical angle defined by $\gamma_1=mn$ and 
the solid line to the Brewster condition $\gamma_1=m\sqrt{n^2+1}$. 
Note that the TE or TM association does not work around the Brewster 
angle.}
\label{fig:MvsReGamma}
\end{figure}

Focusing now on the case of the cylinder (circular cross-section) we 
have the immediate simplification that the Helmholtz equation and
boundary conditions separate in cylindrical coordinates and we can 
look for solutions corresponding to a single component of the angular 
momentum, $m$, instead of the sums in Eq.~\ref{eq:GeneralAnsatz}. In 
the following we will use the following notation:
\be
J_m:=J_m(\gamma_1R)\quad\text{and}\quad H_m:=H^+_m(\gamma_2R)
\ee
where $\gamma_i=\sqrt{n_i^2k^2-k_z^2}, i\in\{1,2\}$ and $R$ is taken 
on the boundary of the domain. With this convention we can write the 
ansatz for the cylinder
\bal
E_z^<(\bm{r};m,j)&=\alpha_mJ_m  (\gamma_1^{m,j}\bm{r})e^{im\varphi}&\bm{r}&<R\\
B_z^<(\bm{r};m,j)&=\xi_mJ_m     (\gamma_1^{m,j}\bm{r})e^{im\varphi}&\bm{r}&<R\\
E_z^>(\bm{r};m,j)&=\upsilon_mH_m^+(\gamma_2^{m,j}\bm{r})e^{im\varphi}&\bm{r}&>R\\
B_z^>(\bm{r};m,j)&=\zeta_mH_m^+ (\gamma_2^{m,j}\bm{r})e^{im\varphi}&\bm{r}&>R.
\label{eq:EzBzcircAnsatz}
\eal
Here, $j$ is the radial mode index enumerating the solutions for a 
given $m$. Using the boundary conditions for the continuity of the 
field, Eqs.~\eqref{eq:mBC1} and \eqref{eq:mBC2}, we get the relations
\be
\upsilon_m=\frac{J_m}{H_m}\alpha_m\quad\quad\zeta_m=\frac{J_m}{H_m}\xi_m\label{eq:outINrelation}.
\ee
Using this, Eqs.~\eqref{eq:mBC3} and \eqref{eq:mBC4} can be rewritten 
in the following form
\begin{widetext}
\be
\mat
{im(n-n^3)\sin\theta J_mH_m}
{\cos^2\!\alpha\, H_m\partial_\rho J_m -n^2\cos^2\!\theta\, 
J_m\partial_\rho H_m}
{n^2\cos^2\!\alpha\, H_m\partial_\rho J_m -n^2\cos^2\!\theta\, 
J_m\partial_\rho H_m}
{im(n^3-n)\sin\theta J_mH_m}
\vect{\alpha_m}{\xi_m}=0\label{eq:agMatrixgeo},
\ee
where the angles are given following the convention in 
Fig.~\ref{fig:3dschematic} B), C), with  $\tan\theta=k_z/\gamma_1$. 
$\theta$ is the interior angle of the ray spiraling up with respect 
to the $(x,y)$ plane and $\alpha$ the corresponding exterior angle. 
In order for this system to have a non-trivial solution the 
determinant needs to vanish, resulting in:
\be
\begin{split}
(1-n^2)^2 m^2\sin^2\!\theta&=\frac{1}{J_mH_m}
\left[\cos^2\!\alpha\, H_m\partial_\rho J_m -\cos^2\!\theta\, 
J_m\partial_\rho H_m\right]\\
&\times\frac{1}{J_mH_m}
\left[\cos^2\!\alpha\, H_m\partial_\rho J_m -n^2\cos^2\!\theta\, 
J_m\partial_\rho H_m\right]\\
&\equiv G^{TM}\cdot G^{TE}.
\end{split}\label{eq:3dresoCond}\ee
\end{widetext}
Where we have defined:
\bal
G^{TE}&=\frac{1}{J_mH_m}\left[\cos^2\!\alpha\, H_m\partial_\rho J_m
-n^2\cos^2\!\theta\, J_m\partial_\rho H_m\right]\nonumber\\
G^{TM}&=\frac{1}{J_mH_m}\left[\cos^2\!\alpha\,H_m\partial_\rho J_m
-\cos^2\!\theta\,J_m\partial_\rho H_m\right].\label{eq:GTM}
\eal
This form is useful since the left-hand side is independent of $k$ 
and vanishes as $\theta \rightarrow 0$ for all $m$, yielding
\bal
0=G^{TE}\cdot G^{TM}&=
\big[H_m\partial_\rho J_m
-n^2J_m\partial_\rho H_m\big]\\
&\times\,
\big[H_m\partial_\rho J_m
-J_m\partial_\rho H_m\big].
\eal
The vanishing of the left bracket describes the resonance condition 
for the usual two-dimensional TE modes and the vanishing of the right 
bracket that for the TM modes~\cite{tureci_thesis,Tureci05}, so one 
recovers the correct limiting behavior as $\theta \rightarrow 0$ 
$(k_z \rightarrow 0)$.  In order to find the resonance wavevectors 
for the spiral modes at $\theta \neq 0$ one needs to find the complex 
roots of Eq.~\eqref{eq:3dresoCond}.  An example of such solutions is 
given in Fig.~\ref{fig:MvsReGamma}, where the roots were found by the 
{\tt SLATEC} routine {\tt dnsqe}~\cite{slatec}.  The series of 
resonances for a given value of $m$ will be labeled by a second 
integer $j$ which indexes the quantized radial momentum in the 
transverse plane.

Further insight into the solutions can be obtained by noting that for 
$k_z=0$ and $m \neq 0$ the TE and TM resonance values differ so that 
at a typical TM resonance, for example, the factor $G^{TE}$ will have 
its typical order of magnitude while the factor $G^{TM}$ vanishes. 
By continuity of these functions with $\theta$ we can expect for 
small $\theta$ that the resonances will have one of the factors 
$G^{TM,TE}$ small while the other is not, and that the corresponding 
resonance will have a TM or TE character, i.e.\ the electric field 
will be predominantly in the $z$-direction or predominantly in the 
$x-y$ plane.  The TE-like resonances will then show a large width 
(imaginary part) near the Brewster angle while the TM-like resonances 
will not.  As the tilt angle $\theta$ increases the spiral modes 
become full mixtures of TE and TM modes and it is no longer possible 
to classify them in this manner.  In Fig.~\ref{fig:MvsReGamma}, 
$\theta$ is small enough to classify them as TE- and TM-like and the 
coloring represents this classification.

Once the resonances are found, the polarization in the far field can 
be determined by rewriting Eq.~\eqref{eq:agMatrixgeo} using the 
functions $G^{TM,TE}$ as
\bal
\xi_m&=i\frac{m\left(n-n^3\right)\sin\theta}{n^2G^{TE}}\alpha_m\\
\alpha_m&=i\frac{m\left(n^3-n\right)\sin\theta}{n^2G^{TM}}\xi_m.
\eal
The coefficients $\alpha_m, \xi_m$ determine the ratio of $E_z$ to 
$B_z$ in the far-field by
\be
P=\frac{B^>_z}{E^>_z}=\frac{\xi_m}{\alpha_m}=
i\frac{m\left(n-n^3\right)\sin\theta}{n^2G^{TE}}.\label{eq:polarization}
\ee
We will however defer detailed analysis of the polarization 
properties of spiral resonances of the cylinder until we have 
developed the theory in a more intuitive semiclassical approximation 
in Sections~\ref{sect:ebk} below.

\subsection{Small $\theta$ expansion }

\begin{figure}
\centering
\psfrag{sinchi}{\hspace{-5mm }\lower1ex\hbox{$\sin\chi=m/nk$}}
\psfrag{alpha}{$\vspace{1cm }\alpha$}
\includegraphics[width=7cm,height=4cm]{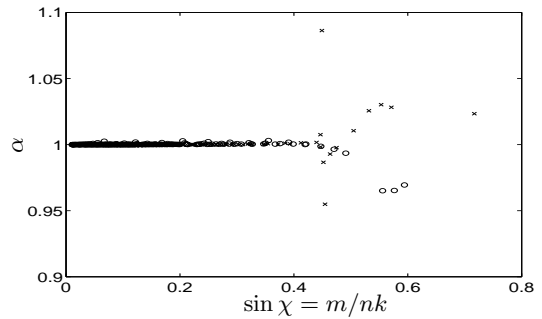}
\caption{Exact calculation of the coefficient for the resonant shift 
(circle) TM-like, (cross) TE-like.  The coefficient is expected to be 
unity for small
$\sin \chi$.}
\label{fig:alphaBessel}
\end{figure}

\begin{figure}[hbt]
\centering
A)\includegraphics[width=5cm]{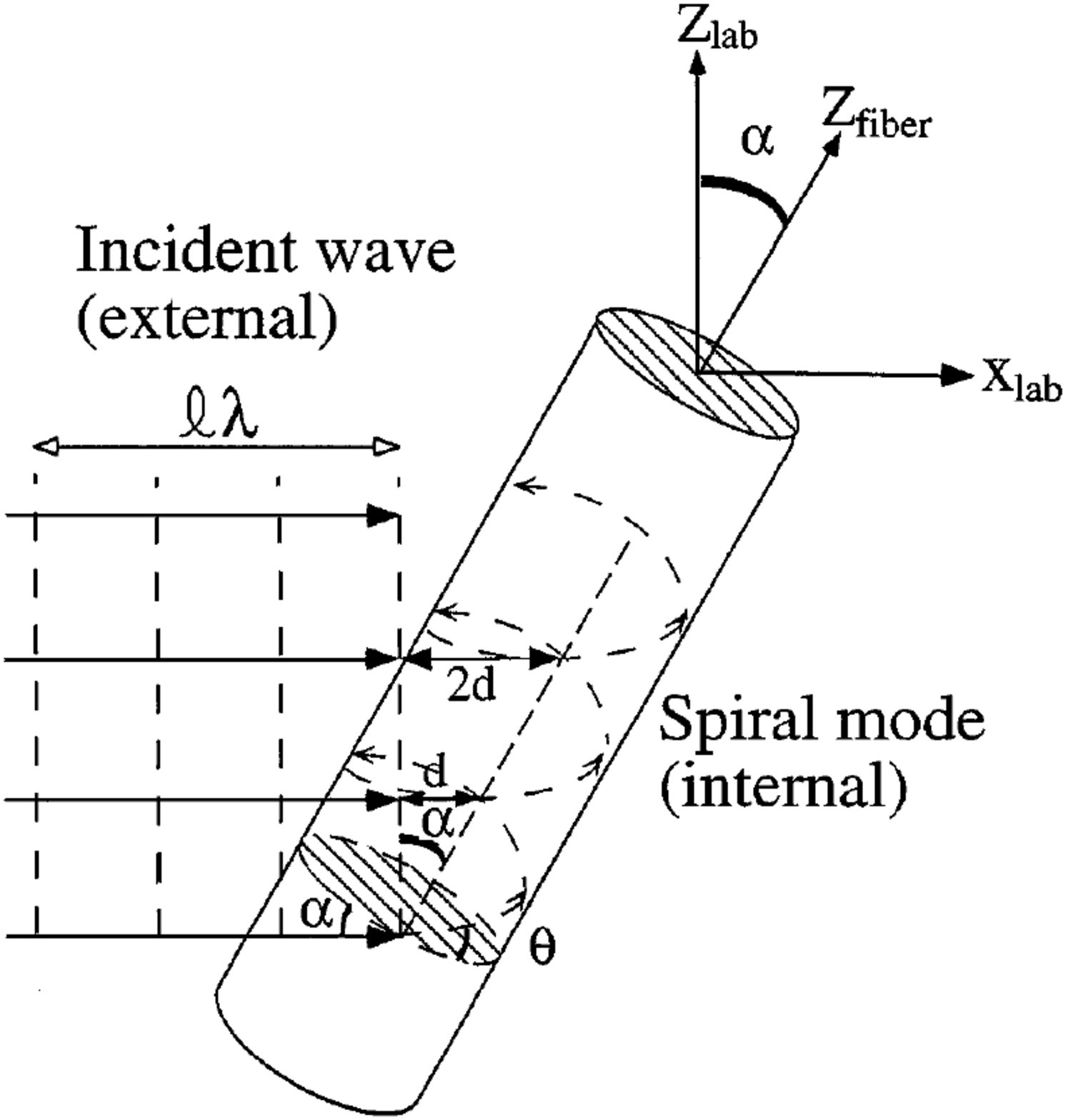}
\psfrag{t}{$\theta$}\psfrag{2pa}{$2\pi a$}\psfrag{2cos}{$2\pi a\cos\theta$}
  B)\includegraphics[width=5cm]{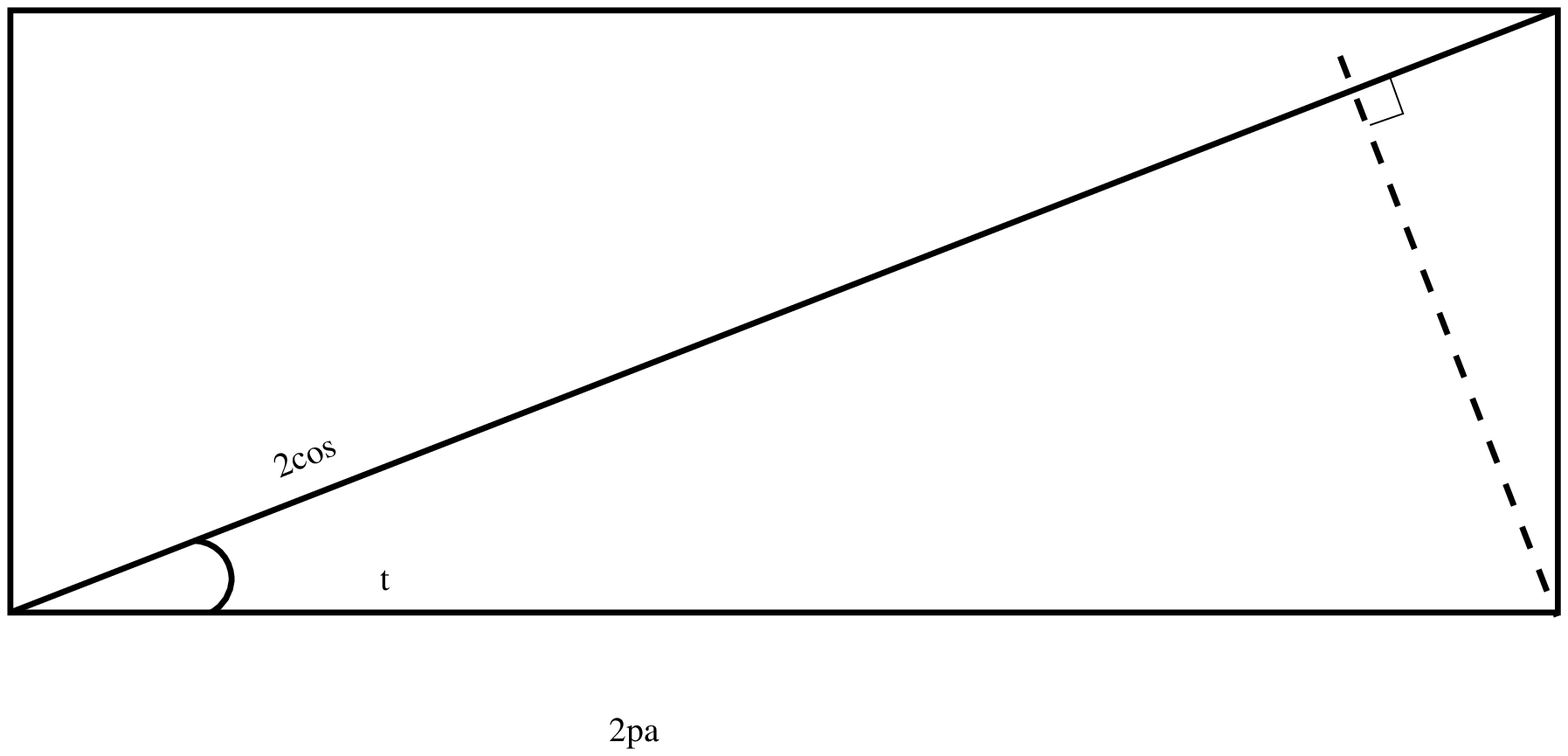}
\caption{A) Schematic of wave-front-matching argument. Internal 
spiral wave of a tilted optical fiber with respect to the incoming 
wave. The phase-matching condition between the spiral mode and the 
external incident wave reduces the effective cavity length for the 
spiral wave by a distance $d/n$, where $n$ is the refractive index. 
B) The spiral quadratic blue shift can be interpreted by unwrapping 
the circular fiber. The dashed lines indicates the wavefront. The 
wavefront matched path is only $2\pi a\cos\theta$, therefore the 
resonances are quadratic blue shifted with the tilt angle. The 
figures are adapted from Poon~\protect\cite{PoonCL98}.}
\label{fig:poon2}
\end{figure}

In the limit of small $k_z$, or $\theta$, we see that 
Eq.~\eqref{eq:3dresoCond} is of order $\theta^2$ on the left hand 
side. Close to a $TE$ or $TM$ like resonance at $k=k_0$,$k_z=0$, one of the factors $G^{TE}$ or $G^{TM}$ is small while the other is not. We expand the 
small term (which vanishes on resonance) to lowest order in $\theta^2$ 
and insist that the resonance condition is satisfied at a slightly shifted value of the resonance wavevector, $nk=n (k_0+\Delta k_0)$. One can then show that 
\bal
\frac{\Delta k_0}{k_0} &=\frac{1}{2}\alpha\theta^2\label{eq:resonantApprox}
\eal
where $\alpha$ is given by a ratio of Bessel functions, which is 
plotted in Fig.~\ref{fig:alphaBessel}.  Note that the coefficient 
$\alpha \approx 1$ for relatively small $\sin \chi$; exactly this 
value follows from the EBK quantization formula near normal 
incidence discussed in Section~\ref{sect:ebk}.
This result can be compared to experiments done by Andrew 
Poon~\cite{PoonCL98}, where a tilted optical glass 
fiber was illuminated with an unfocused Gaussian beam.
A simple ``wavefront matching'' argument for this blue shift was 
given in Ref.~\refcite{PoonCL98} and is reproduced in 
Fig.~\ref{fig:poon2}. An incident plane wave propagates along the 
$X_\text{lab}$ direction and is incident onto a rod tilted by an 
angle $\alpha$. An upward propagating spiral resonance requires that 
the spiral wave be in phase with the incident wave farther up the 
tilted fiber. However upper incident wave of the same phase front 
must travel an extra distance $d$ to reach the fiber when it is 
tilted. Thus the phase-matching condition between the internal wave 
and the extended incident wave reduces the effective cavity length 
for the spiral wave by a distance $d/n$, where $n$ is the refractive 
index. Fixing the quantum numbers of the resonances, this implies a 
quadratic blue-shift as a function of the tilt angle $\theta$.

\begin{figure}[hbt]
\centering
\psfrag{nk}{$nk$}\psfrag{theta}{$\theta$}
\includegraphics[width=7cm, height=4cm]{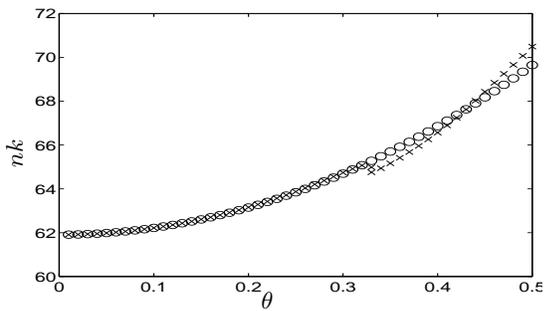}
\caption{Shift of the resonance condition with the quantum numbers 
$j=21,m=20$ ($\sin\chi\approx0.33$) and $n=2$ with respect to the 
internal tilt angle $\theta$. Crosses are the numerical solutions 
following Eq.~\eqref{eq:angQuant}, and circles following the small 
$\theta$ expansion Eq.~\eqref{eq:resonantApprox}. Note the 
discontinuity in the exact numerical values around $\theta=0.33$; 
this is due to the 
onset of the polarization critical angle which 
will be discussed in Section~\protect\ref{sect:pca }. This 
discontinuity occurs beyond the regime of the small $\theta$ 
expansion.}
\label{fig:blueShift}
\end{figure}

In Fig.~\ref{fig:blueShift} we compare the blueshift obtained from 
the exact numerical solution of Eq.~\eqref{eq:angQuant} to the small 
$\theta$ expansion Eq.~\eqref{eq:resonantApprox}. The agreement is 
quite good. We will see at the end of the next section that the EBK 
quantization will give the same quadratic blue shift.

\section{The EBK quantization conditions}\label{sect:ebk}
As already noted, using the quantized solutions of resonance 
condition Eq.~\eqref{eq:3dresoCond}, we can easily calculate the 
polarization in the farfield. However, to get insight into the 
dependence of the polarization on the internal ray motion, a more 
appealing approach is to use the eikonal method, and the 
Einstein-Brillouin-Keller (EBK) type formulation of the resonance 
conditions. The eikonal method in general refers to finding 
approximate solution of the wave equation with a specific ansatz 
which is expected to be good in the short-wavelength limit ($k 
\rightarrow \infty$). The EBK method describes how to apply that 
ansatz to boundary value problems, typically assuming Dirichlet or 
Neumann boundary conditions~\cite{Keller58}.  The approach has been 
used for example to find approximate quantization formulas for the 
circular and elliptical billiard systems~\cite{Keller60}. Recently it 
was generalized to treat the scalar Helmholtz equation for a 
dielectric billiard in two dimensions~\cite{tureci_thesis}. However 
it was also emphasized that the EBK method only works for the small 
subset of boundary shapes for which the ray motion within the 
boundary is integrable~\cite{Tureci05}, a point which goes all the 
way back to Einstein's original paper in 1917~\cite{einstein17}.  The 
motion of a ray within an infinite cylinder is also integrable (the 
energy and $z$-components of linear and angular momentum are 
conserved) and a generalization of the EBK method should work in this 
case also. The necessary generalization is to introduce the boundary 
conditions appropriate for the coupled $E_z$ and $B_z$ components of 
the field; this will lead to a generalization of the Fresnel 
coefficients for a plane interface.

We study the vector Helmholtz Equation~\eqref{eq:EzBzHelmholz} for 
$E_z(x,y),B_z(x,y)$ in the {\em semiclassical limit} $k,\gamma_{1,2} 
\rightarrow\infty$; in this limit we expect the solutions to have 
rapid phase variations and relatively slow amplitude variations. The 
generalized EBK ansatz for the quasi-bound solutions of the vector 
Helmholtz Eq.~\eqref{eq:EzBzHelmholz} can be written as
\be
\vect{E_z}{B_z}=\Psi({\bm r})=A_1e^{i\gamma S_1({\bm 
r})}+A_2e^{i\gamma S_2({\bm r})}
\ee
where $A_{1,2}$ are two-component vectors and $S$ is the Eikonal. 
Note that all the functions are defined on the two-dimensional $x-y$ 
plane. Following Refs~\refcite{Keller60,tureci_thesis} we can write 
the general quantization condition
\be
\gamma \oint_{\Gamma_i} d{\bm q \cdot \nabla}S=2\pi l_i + \Phi_i 
\quad i=1,2.\label{eq:EBKint}
\ee
Here the quantity $\nabla S$ is the gradient of the phase functions 
$S_1,S_2$ considered as the two sheets of a double valued vector 
field defined on the cross-section, $l_i$ are integers and $\Gamma_i$ 
refer to topologically irreducible set of loops. To avoid confusion 
in this section we temporarily drop the subscript $\gamma_1 
\rightarrow \gamma$.  Keller showed that in order for the EBK 
solution to be single-valued it is necessary that these loop 
integrals of the phase be quantized and that {\it any} two 
topologically inequivalent and non-trivial loops can be chosen. 
$\Phi_i$ is a total phase shift due to caustics and boundary 
scattering; for the scalar two-dimensional Helmholtz equation in a 
circle these phase shifts are known for Dirichlet, Neumann and 
dielectric boundary conditions.  For the case of a dielectric circle 
the phase shifts are complex in general, representing either 
refraction out of the circle or the phase shift at the boundary due 
to total internal reflection~\cite{tureci_thesis}.  In order to get 
the appropriate ray dynamics for the spiral modes it is easily shown 
that the eikonals $S_1,S_2$ must be identical to those of the 
two-dimensional circular billiard (i.e. $k_z=0$); the new feature 
here is an additional eigenvalue condition on the amplitude 
two-vector, which will determine the eigenpolarization directions. 
This will lead to a modification of the phase shifts $\Phi_i$ with 
respect to the two-dimensional case.

\begin{figure}[hbt]
\centering
\psfrag{C1}{${\cal C}_1$}\psfrag{C2}{${\cal C}_2$}
\psfrag{sin}{$\sin\chi$}\psfrag{G1}{$\Gamma_1$}
A)\includegraphics[width=5cm]{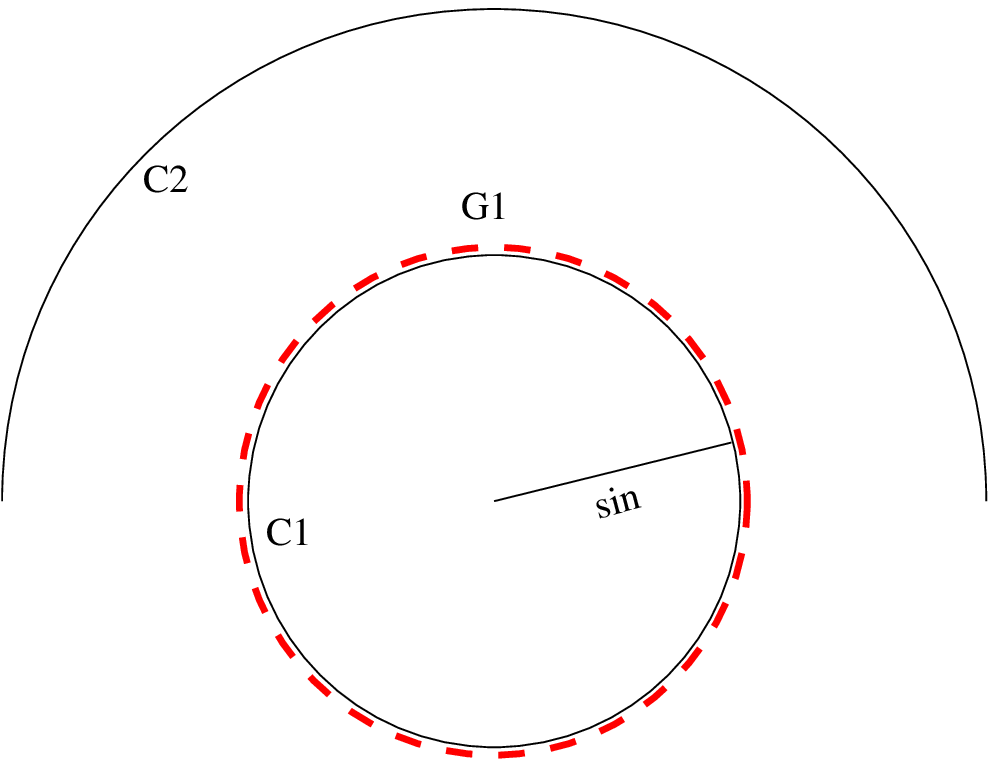}
\psfrag{cos}{$\cos\chi$}
\psfrag{c}{$\chi$}\psfrag{a}{$\alpha$}\psfrag{G2}{$\Gamma_2$}
B)\includegraphics[width=5cm]{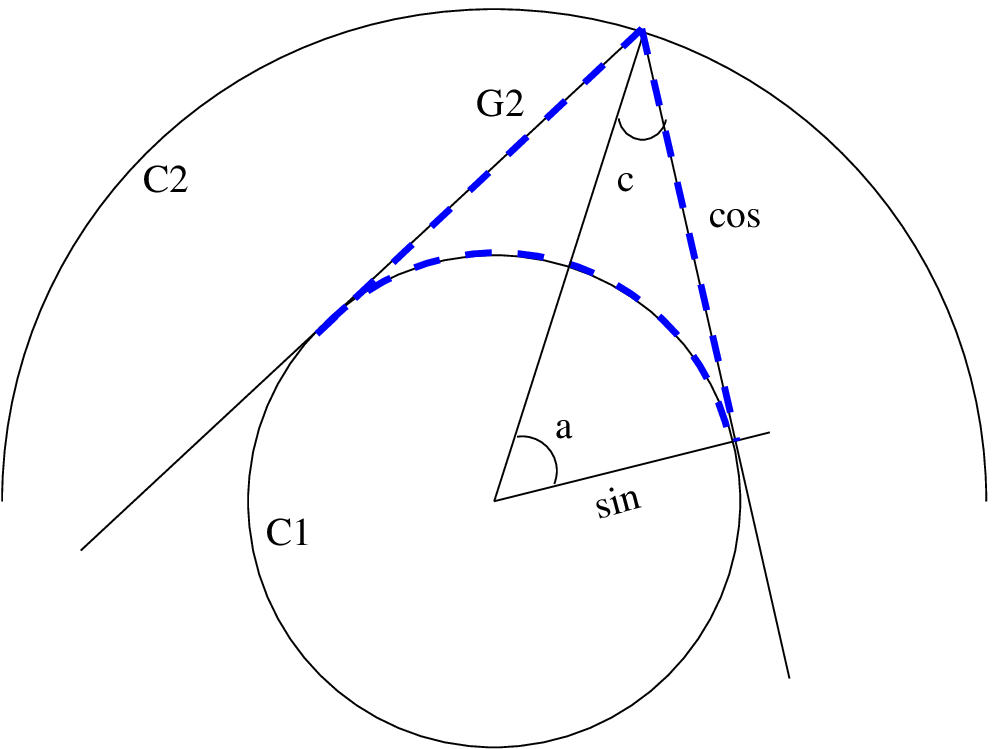}
\caption{A) Path of the first curve $\Gamma_1$. B) Second path 
$\Gamma_2$, of length L.}
\label{fig:EBKlength}
\end{figure}

Before discussing the latter point we briefly review the quantization 
relations assuming the $\Phi_i$ are known. Two conventional loops for 
implementing the quantization conditions are shown in 
Fig.~\ref{fig:EBKlength}.  The first loop, $\Gamma_1$, goes just 
outside the inner turning point of a ray of fixed angular momentum; 
$\nabla S$ points in the direction of the ray so this integral just 
yields the length of the caustic for this ray.  No caustic surfaces 
are crossed and the path does not touch the boundary, so the phase 
$\Phi_1=0$.  Thus the first loop gives a relation equivalent to 
angular momentum conservation,
\be
\sin\chi=\frac{m}{\gamma R}\label{eq:angQuant}
\ee
where we have replaced the integer $l_1 \rightarrow m$ to conform to 
our earlier notation.

The second loop $\Gamma_2$ gives the quantization condition for the 
reduced wavevector $\gamma$ in terms of the path length $L$ of the 
loop (the vector field $\nabla S$ is everywhere parallel to this path)
\bal
\gamma L&= 
\left[2\cos\chi-2\left(\frac{\pi}{2}-\chi\right)\sin\chi\right] 
R\notag\\
  &= 2 \pi j + \Phi_2,\label{eq:EBKquant2}
\eal
where we have replaced the integer $l_2$ by $j$. This is exactly the 
relation we would find for the two-dimensional billiard problem, 
except that the transverse wavevector $\gamma =\sqrt{n^2k^2-k_z^2}$ 
has replaced the full wavevector $nk$ and the appropriate phase shift 
$\Phi_2$ needs to be determined, and will differ from the 
two-dimensional case.

\subsection{Semi-classical boundary conditions for dielectric 
rod}
\begin{figure}[hbt]
\centering
\psfrag{g1}{$\gamma_1$}\psfrag{g2}{$\gamma_2$}
\psfrag{psiI}{$\Psi^i$}\psfrag{psiO}{$\Psi^t$}\psfrag{psiR}{$\Psi^r$}
\includegraphics[width=6.5cm,clip]{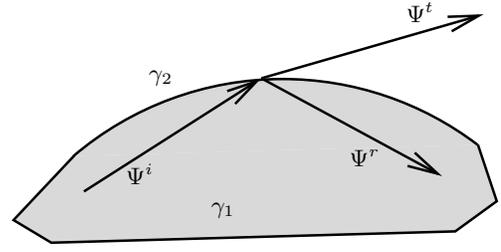}
\caption{Schematics of scattering on the projected plane. We will 
expand the wave solution inside into an incoming component $\Psi^i$ 
and a reflected $\Psi^r$. The outside transmitted component is given 
by $\Psi^t$.}
\label{fig:schematicEBK}
\end{figure}

\index{EBK!Ansatz}\label{sect:ebkAnsatz} \index{EBK!Ansatz}
To treat the spiral modes of a dielectric cylinder in this approach 
we need to project the three-dimensional boundary conditions 
corresponding to Snell and Fresnel's laws into two dimensions.  The 
relevant angles for this projection are all defined in 
Fig.~\ref{fig:3dschematic} Since we are in the ray optics limit we 
can regard the ``scattering" of the eikonal from the boundary as the 
scattering of a plane wave from the tangent plane.  Because of our 
assumption of only outgoing waves, it is sufficient to assume only an 
incident, reflected and a transmitted wave (see 
Fig.~\ref{fig:schematicEBK}).

\subsubsection{Generalized Snell's law}
\begin{figure}[hbt]
\centering
\psfrag{theta}{\lower1ex\hbox{$\sin\theta$}}\psfrag{f}{\hspace{-2ex}$f(\theta)$}
\includegraphics[width=7cm,height=4cm]{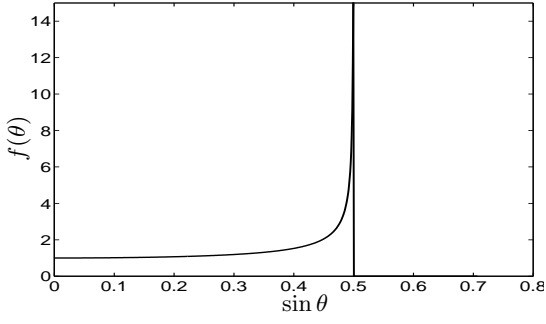}
\caption{Functional dependence of $f(\theta)$ for $n=2$. Note that 
$f(\theta)$ diverges at the critical angle when $\sin\theta=1/n$.}
\label{fig:fSnell}
\end{figure}

The form of each of the fields is given by
\be
\Psi^m=\vect{E_{z,m}}{B_{z,m}}e^{i\gamma{S}_m},\With m\in\{i,r,t\}
\ee
The gradient of the Eikonal $\nabla\!{S}$ gives the direction of the 
the ray and is of constant length $|\nabla\!{S}|=n$.
We find
\be
\begin{array}{r@{=}r@{\hspace{5mm}}r@{=}r@{\hspace{5mm}}r@{=}l}
\partial_n\bm{S}^i&i\gamma_1\cos\chi&
\partial_n\bm{S}^r&-i\gamma_1\cos\chi&
\partial_n\bm{S}^t&i\gamma_2\cos\sigma\\
\partial_t\bm{S}^i&i\gamma_1\sin\chi&
\partial_t\bm{S}^r&i\gamma_1\sin\chi&
\partial_t\bm{S}^t&i\gamma_2\sin\sigma\end{array}\label{eq:Scomponents}.
\ee
The first set of boundary conditions, the continuity of the field 
across the boundary, Eqs.~\eqref{eq:mBC1} and \eqref{eq:mBC2} becomes:
\be
\vect{E_z}{B_z}^ie^{i\gamma_1{S}^i}+
\vect{E_z}{B_z}^re^{i\gamma_1{S}^r}=
\vect{E_z}{B_z}^te^{i\gamma_2{S}^t}\label{eq:EzBzcont}.
\ee
Since these equations need to hold everywhere on the boundary, the 
phases need to be equal, thus yielding
\be
\gamma_1S^i=\gamma_1S^r=\gamma_2S^t.
\ee
Using the fact that the tangent components are continuous, we obtain
\be
\gamma_1\sin\chi=\gamma_2\sin\sigma\quad\Rightarrow\quad 
\sin\chi=\frac{\gamma_2}{\gamma_1}\sin\sigma.\label{eq:generalSnell}
\ee
This equation can be identified as the projection of {\em Snell's 
Law}\index{Snell's Law!general derivation EBK}  into the transverse 
plane.  Note that when $k_z=0$ we recover the usual result
\be
n\sin\chi=\sin\sigma\label{eq:SnellsLaw}.
\ee
We can write the projected Snell's Law in a completely geometric 
fashion noting $\gamma_1/\gamma_2 = \tan \theta/ \tan \alpha$,
\be
\sin\sigma=\frac{\gamma_1}{\gamma_2}\sin\chi=\frac{\sin\alpha\,\cos\theta}{\cos\alpha\,\sin\theta}\sin\chi=f(\theta)n\sin\chi.
\ee
With the function $f(\theta)$ given by
\be
f(\theta)=\frac{\cos\theta}{\sqrt{1-n^2\sin^2\theta}}=\sqrt{\frac{1-\sin^2\theta}{{1-n^2\sin^2\theta}}}\ge1.
\ee
Hence, as is clear geometrically, the projected Snell's law, leads to 
total internal reflection of the projected motion before the critical 
angle $\sin \chi = 1/n$ is reached in the plane (this is simply 
because the actual angle of incidence is steeper than the projected 
angle due to the
$z$-motion); the function $f(\theta)$ which determines the effective 
critical angle is plotted in Fig.~\ref{fig:fSnell}.

\subsubsection{Generalized Fresnel Matrices}\index{Fresnel 
coefficients!general derivation}\label{sect:genFresnel}
The kinematics of the projected ray motion has been determined above 
simply from the continuity of the tangential components of the 
fields; the transport of ray flux across the boundary will now be 
determined from the normal derivative boundary conditions.
\be
\mf{B}^i\vect{E_z}{B_z}^ie^{i\gamma_1{S}^i}+
\mf{B}^r\vect{E_z}{B_z}^re^{i\gamma_1{S}^r}=
\mf{B}^t\vect{E_z}{B_z}^te^{i\gamma_2{S}^t}\label{eq:EzBzMat}
\ee
where the matrices $\mf{B}$ are derived from the boundary conditions 
Eqs.~\eqref{eq:mBC3} and \eqref{eq:mBC4} and given by the matrices
\bal
\mf{B}^{(i,r)}&
=\mat{(n-n^3)\sin\theta\cdot\partial_t}{\cos^2\alpha\cdot\partial_n}
{n^2\cos^2\alpha\cdot\partial_t}{(n^3-n)\sin\theta\cdot\partial_n}\label{eq:BCmatrix}\\
\mf{B}^t&=\mat{0}{n^2\cos^2\theta\cdot\partial_n}
{n^2\cos^2\theta\cdot\partial_n}{0}.
\eal
Here $\partial_n$ and $\partial_t$ are the normal and tangential 
derivatives. We can now relate the incoming field to the outgoing and 
the reflected using the boundary conditions Eq.~\eqref{eq:EzBzcont} 
and Eq.~\eqref{eq:EzBzMat}
\bal
\Psi^r&=R\Psi^i\label{eq:defRmatrix}\\
\Psi^t&=T\Psi^i\label{eq:defTmatrix}
\eal
where $R$ and $T$ are the general Fresnel matrices given by
\bal
R&=\left(\mf{B}^r-\mf{B}^t\right)^{-1}\left(\mf{B}^t-\mf{B}^i\right)\label{eq:Rmatrix}
\intertext{and similarly}
T&=\left(\mf{B}^t-\mf{B}^r\right)^{-1}\left(\mf{B}^i-\mf{B}^r\right).
\eal
In the limit $\theta \rightarrow 0 \Rightarrow \alpha \rightarrow 0$ 
these $2 \times 2$ matrices become diagonal and take the form
\bal
R&=\mat{\frac{n\cos\chi-\cos\sigma}{n\cos\chi+\cos\sigma}}
{0}{0}{\frac{\cos\chi-n\cos\sigma}{\cos\chi+n\cos\sigma}}
&\widehat{=}&& R&=\mat{r_s}{0}{0}{-r_p}\label{eq:reflectionMatrix}
\\
T&=\mat{\frac{2n\cos\chi}{n\cos\chi+\cos\sigma}}
{0}{0}{\frac{2\cos\chi}{\cos\chi+n\cos\sigma}}
&\widehat{=} &&T&=\mat{t_s}{0}{0}{t_p}.
\eal
We recognize the diagonal elements as the Fresnel coefficients for TM 
(denoted by subscript $s$) and TE (subscript $p$) plane waves 
incident on a dielectric interface.  The diagonal nature of the 
matrices implies that for $\theta = 0$ these polarization states are 
preserved.  When $\theta \neq 0$ the matrices have off-diagonal 
elements implying the mixing of polarization states upon reflection
(strictly speaking these matrices mix $E_z$ and $B_z$ upon 
reflection, which will be shown to be equivalent to rotating the 
local polarization).
Recall that we are applying the EBK quantization 
condition~\eqref{eq:EBKint} to the two-vector $(E_z(x,y),B_z(x,y))$; 
when we integrate around the path B in Fig.~\ref{fig:EBKlength} which 
touches the boundary, we must impose the correct dielectric boundary 
conditions on this two-vector.  In general this changes the ratio of 
$E_z$ to $B_z$ and will lead to a multi-valued solution as we 
complete the closed loop (for uniform index rods, only boundary 
scattering leads to the rotation of the two-vector).  Therefore in 
order to have a single-valued solution the ratio of $E_z$ to $B_z$ 
must be unchanged upon reflection, i.e.\ the two-vector $\Psi_i$ must 
be an eigenvector  $\mf{a}$ of the reflection matrix,
\be
R\mf{a}=\Lambda\mf{a} \label{eq:EigPolEBK}.
\ee
We thus see that for the spiral modes of the cylinder there are two 
allowed mixtures of TM and TE polarization for each resonance 
labelled by angular momentum $m$ and wavevectors $\gamma_1,k_z$; the 
eigenvalues of the $R$ matrix $\Lambda = e^{i\eta}$  will give the 
extra phase  shift $\Phi_2$ needed to complete the EBK quantization 
condition in Eq.~\eqref{eq:EBKquant2} ($\Phi_2 = \eta + \pi/2$ where 
the term $\pi/2$ comes from the caustic phase shift).
As already shown above, at $\theta = 0$ the $R$ matrix is diagonal, 
conventional TE and TM states are eigenvectors and the eigenvalues 
are just the Fresnel reflection coefficients, $r_s,r_p$.  These have 
the familiar property of being purely real and less than unity for 
incident angle $\chi$ below total internal reflection, and complex 
numbers of modulus unity for angles above TIR.  Hence one has pure 
reflection and refraction with no phase shift below TIR, and  pure 
phase shift and no refracted wave above TIR.  The phase shifts for 
totally internally reflected TM and TE waves are different and 
well-known functions of $\chi$ \cite{jackson_book}.  This familiar 
behavior is modified for the spiral modes.

\begin{figure}[t!]
\centering
\psfrag{eigen}{$|\Lambda|$}
\psfrag{sinchi}{\lower1ex\hbox{$\sin\chi$}}
\includegraphics[width=7cm,height=4cm]{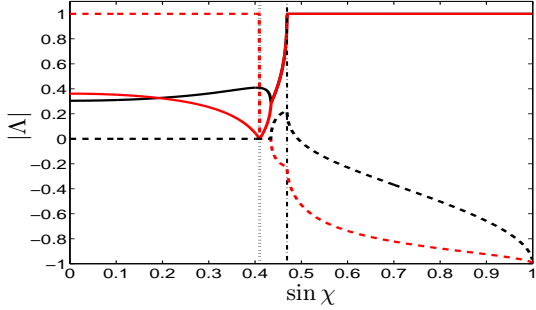}
\caption{Absolute value (solid red and black lines) and the phase 
divided scaled by $\pi$ (dashed red and black lines) of the 
eigenvalues of $R$ for $n=2$ and $\tan\theta=0.2$ vs.\ $\sin\chi$. 
The dotted black horizontal line is the Brewster angle and the black 
dashed line is the critical angle. Red indicates the TE-like 
component and blue the TM.}
\label{fig:EBKeignevalueN2T0.2}
\end{figure}

The behaviour of the eigenvalues $\Lambda=e^{i\eta}$ of the matrix 
$R$ for $k_z\neq0$ is shown as a function of the angle 
$\sin\chi=m/\gamma$  in Fig.~\ref{fig:EBKeignevalueN2T0.2}. Note that 
$\eta$ is in general complex, and we are plotting here its magnitude. 
The magnitude of the eigenvalues are different up to a point between 
the Brewster angle (vertical dashed), and the CA (vertical dashed). 
At this point the the eigenvalues become complex conjugates of each 
other. We will call this point, the {\em polarization critical angle} 
(PCA) and will explain how it is related to the far-field 
polarization below. Hence spiral modes acquire a phase shift upon 
reflection before they reach the critical angle. Not until the CA do 
$\Lambda_{1,2}$ lie on the complex unit circle as they should for 
TIR. Thus we have a new phenomenon, a phase shift for  a refracted 
wave.
Having determined the two possible values of $\Phi_2$, we can write 
the quantization condition for the spiral resonances as a 
transcendental equation
\be
2\gamma\sqrt{1-\frac{m^2}{\gamma^2}}+2m\arcsin\left(\frac{m}{\gamma}\right)=
2\pi j + \frac{\pi}{2} + m\pi +\zeta+i\ln|r|,
\ee
where $j,m$ are integers, $\Lambda=re^{i\zeta}$, and we have used the 
angular momentum quantization condition Eq.~\eqref{eq:angQuant}. We 
can simplify this result further to get an explicit solution for rays 
near normal incidence in the plane, so that $\sin \chi \rightarrow 0$:
\be
\gamma = \pi\left[j+\frac{m}{2}+\frac{1}{4}\right] 
+f(\chi,\theta)\label{eq:planeapprox }
\ee
where the contributions of the phase shift and the loss at the 
boundary have been combined in $f(\chi)$. We will analyze the 
function $f(\chi,\theta)$ further in the next section on polarization 
properties of the spiral modes.

 From Eq.~\eqref{eq:planeapprox } we can derive the blue shift for 
small $\theta$ by noting the the right hand side without 
$f(\chi,\theta)$ is just the resonant condition for the circle. From 
the definition of $\gamma=\sqrt{nk^2-k_z^2 }=nk(1-\sin^2\theta)$ we 
can write
\be
nk=nk_o\left(1+\frac{1}{2}\theta^2\right),
\ee
where $nk_o$ is the resonance condition for the circle $\theta=0$. 
Comparing to Eq.~\eqref{eq:resonantApprox} above we see that this implies that
the 
coefficient of the small $\theta$ quadratic blue-shift should be 
$\alpha = 1$ for small $\sin \chi$, just as we found in
Fig.~\ref{fig:alphaBessel}
above.  As noted there, this coefficient changes slightly when the 
polarization critical angle is reached; this change is
captured by 
the contribution from $f(\chi,\theta)$ we have just neglected.
In Tab.~\ref{tab:ebk3d} we compare the resonances found by the exact 
wave matching method (Eq.~\eqref{eq:3dresoCond}) and by the EBK 
method finding good agreement for $\theta =0.1,0.2$.

%{\tt we still should explicitly analyze the region where the eigenvalues are complex conjugates as there we expect to have a new correction to the resonance condition}

\begin{table}\small\centering
\caption{spiral resonances of the cylinder with $n=2,\theta=0.1, 
0.2$. We compare the resonances calculated by the solution of 
Eq.~(\ref{eq:3dresoCond}) and the EBK method, finding good agreement. 
Although resonances can no longer be classified as TM or TE, 
classification as TM-like or TE-like is based on which factor 
$G^{TE,TM}$ is small at the resonance as discussed 
above}\label{tab:ebk3d}\vspace{2ex}
\begin{tabular}{r|c|r|cr}
\multicolumn{2}{c|}{$\theta=0.1$}&\multicolumn{1}{c}{ 
exact}&\multicolumn{2}{c}{EBK}\\
m & &\multicolumn{1}{|c|}{kR}&j&\multicolumn{1}{c}{kR}\\\hline
18& TE&    100.52083-0.27170i    &55&100.520468-0.271695i\\
20& TE&    100.42571-0.27098i    &54&100.425346-0.270979i\\
44& TE&    100.06672-0.25117i    &43&100.066168-0.255107i\\
74& TE&    100.30341-0.20475i    &31&100.301095-0.204629i\\
98& TE&    101.41861-0.09114i    &23&101.382400-0.075236i\\\hline
\multicolumn{2}{c|}{$\theta=0.2$}&&&\\\hline
  5 & TM&  99.62235-0.59423i  &    29&  99.61990-0.59421i \\
17 & TE&  99.86057-0.55418i  &    23&  99.86699-0.55433i \\
20 & TM&  99.29851-0.56992i  &    22&  99.29473-0.56984i \\
34 & TE& 100.17837-0.84848i  &    16& 100.20152-0.85155i \\
39 & TE&  99.78148-1.35232i  &    14&  99.74088-1.36053i \\
57 & TE&  99.76787-0.00005i  &    8 &  96.79993+0.00009i \\
62 & TM& 100.78974-0.00000i  &    7 & 100.86339+0.00000i \\
\end{tabular}
\end{table}
In general we can always calculate the $R$ matrix via 
Eq.~\eqref{eq:Rmatrix}, find the eigen-polarization directions, and 
subsequently act with $T$ on the internal eigen-polarizations to 
determine the corresponding polarization in the far-field. A 
physically more transparent method to do this involves the a 
reformulation of the problem in terms of the actual polarization 
vector, rather than the two-vector $(E_z,B_z)$. Below, we will 
introduce an equivalent matrix description in terms of the Jones 
Algebra which we will subsequently show to be exactly equivalent to 
the $(E_z,B_z)$ description.

\section{Jones formulation of polarization 
properties}\label{sect:jones}
\begin{figure}[!t]
\centering
\psfrag{sinchi}{\lower1ex\hbox{$\sin\chi$}}\psfrag{eigen}{$\nu_{1,2}$}
\includegraphics[width=7cm, height=4cm]{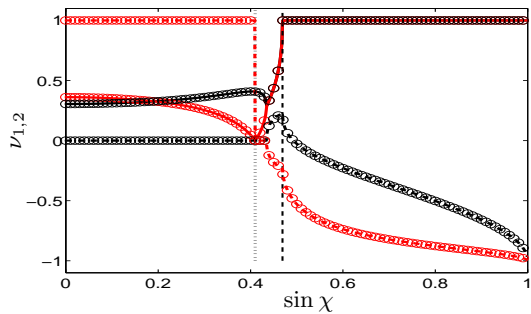}
\caption{Comparison of the of the eigenvalues of the rotated Jones 
matrix, absolute value (solid line red/black) and the phase (divided 
by $\pi$) (dashed line red/black), to the eigenvalues of $R$ 
(red/black circles). Parameters of the calculation are 
$\tan\theta=0.2$ and $n=2$. }
\label{fig:eigenJones}
\end{figure}

\index{Jones matrices}
We introduce the parallel and perpendicular components of the 
electric field, ${E_p}$ and ${E_s}$ in the local coordinate system 
defined by the plane of incidence.  The Jones vector~\cite{jones41} 
which describes the local polarization is
\index{Jones Algebra!Jones vector}
\be
\bm{E}=\vect{E_p}{E_s}=\vect{E_{0p}e^{i\phi_p}}{E_{0s}e^{i\phi_s}}
\ee
with $E_{0s}$ and $E_{0p}$ being the magnitude of the electric field 
and the phases $\phi_s$ and $\phi_p$. The Jones matrix for the 
reflection and transmission at a dielectric interface is give by
\index{Jones Algebra!Jones matrix}
\bal
J_r&=\mat{-r_p}{0}{0}{r_s}\quad\mbox{ (for reflection), }\\
J_t&=\mat{t_p}{0}{0}{t_s}\quad\mbox{ (for transmission)}
\eal
and would describe any series of reflection in the same plane of 
incidence.  However a spiraling ray in the cylinder changes its plane 
of incidence at each reflection, so that we need to rotate our 
coordinate system into the new plane of incidence between each 
reflection before we apply the Jones reflection and transmission 
matrices.  (It is precisely this rotation of the plane of incidence 
for a spiraling ray which is the physical reason behind the 
non-conservation of TE or TM polarization). If the angle between the 
two planes is given by $\xi$ we can do this by multiplying the Jones 
vector by the rotation matrix $\mf{R}(\xi)$. This angle $\xi$ can be 
determined for a general cylindrical symmetry~\cite{schwefel_thesis}. 
Due to the rotational and translational symmetry of the cylinder, the 
polarization at each angle $\phi$ on the cylinder must be the same 
for any value of $z$ and must be described by the same Jones vector 
when referred to the plane of incidence at that point.  Therefore the 
Jones vector emerging from a reflection must be the same as the 
incident Jones vector once the coordinate system has been rotated 
into the new plane of incidence.  This yields an eigenvalue condition 
for the Jones vectors describing the spiral modes of the cylinder,
\be
\mf{J}\vect{E_p}{E_s}_{1,2} = \nu_{1,2}\vect{E_p}{E_s}_{1,2}
\ee
where $\mf{J}=\mf{R}(\xi)J_r$. Thus the two polarization states are 
the two eigenvectors of $\mf{J}$ and their eigenvalues $\nu_{1,2}$ 
describe the phase shifts and refractive losses at each reflection, 
just as do the eigenvalues of $R$ for the two-vector $E_z,B_z$ 
studied earlier.  Therefore, the two matrices $\mf{J}$ and $R$ are 
related by a similarity transformation and have the same 
eigenvalues~\cite{schwefel_thesis} which is confirmed in 
Fig.~\ref{fig:eigenJones}.

\section{Polarization critical angle}\label{sect:pca }
Realizing that the eigenvalues of interest are obtained from the 
product a rotation and a diagonal Jones matrix with known entries 
allows us to understand the behavior of the these eigenvalues rather 
simply.  The eigenvalues can now be written in terms of the rotation 
angle of the plane of incidence, $\xi$ and the Fresnel reflection 
coefficients $r_s,r_p$,
\be
\nu_{1,2}=\frac{1}{2}\cos\xi\left(r_s-r_p\right)\pm
\frac{1}{2}\sqrt{\cos^2\!\xi\,\left(r_s-r_p\right)^2 +4r_pr_s}\label{eq:eigenJones}.
\ee
\begin{figure}[!t]
\centering
\psfrag{sinchi}{\lower1ex\hbox{$\sin\chi$}}\psfrag{eigen}{$\nu_{1,2}$}
\psfrag{theta}{$\theta$}
A)\includegraphics[width=7cm,height=4cm]{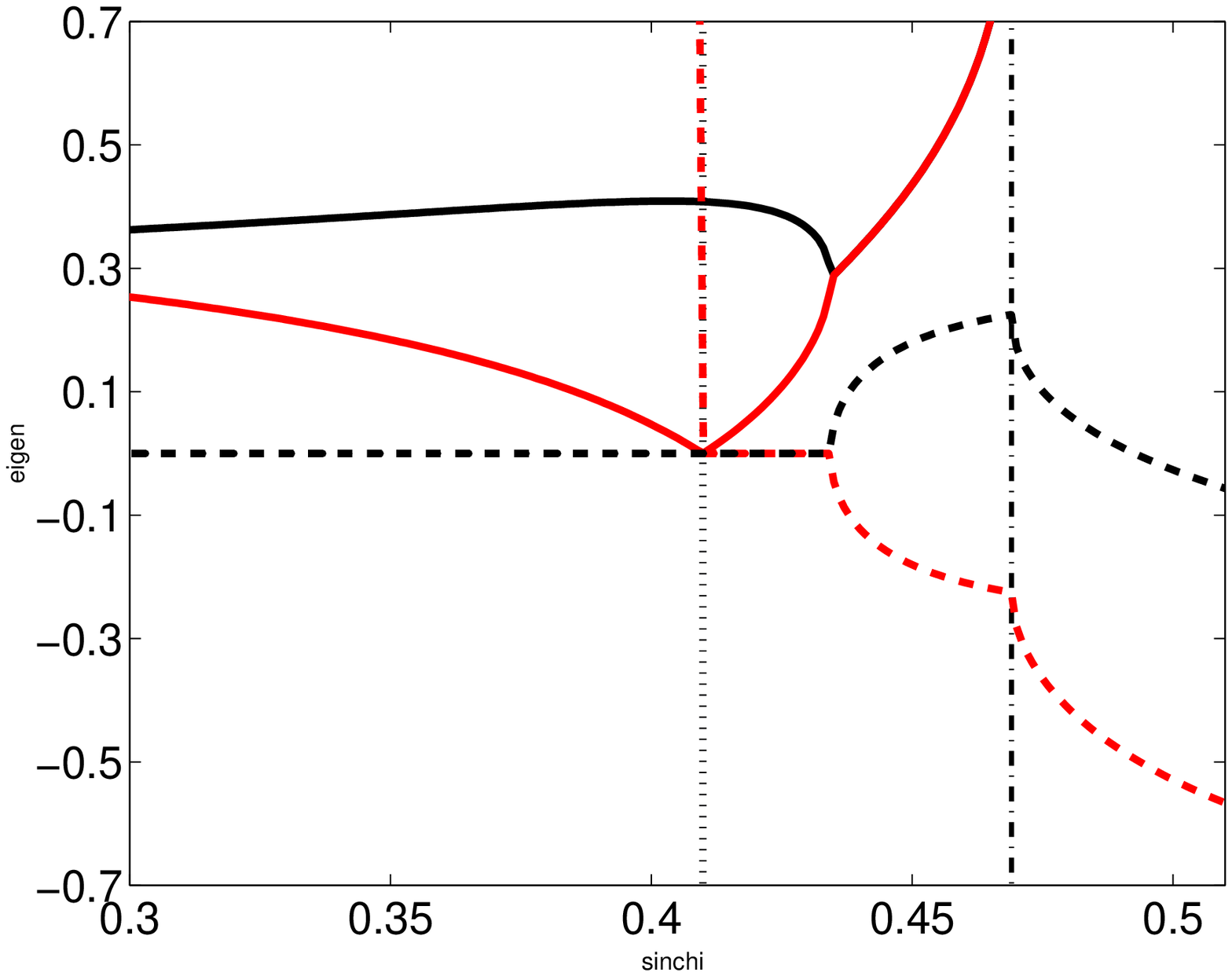}
\psfrag{sin( )}{\raise1ex\hbox{$\sin\chi$}}
\psfrag{theta}{\lower1ex\hbox{$\theta$}}
%B)\includegraphics[width=7.5cm,height=4cm]{newestCriticalAngle1.eps}
B)\includegraphics[width=7cm,height=4cm]{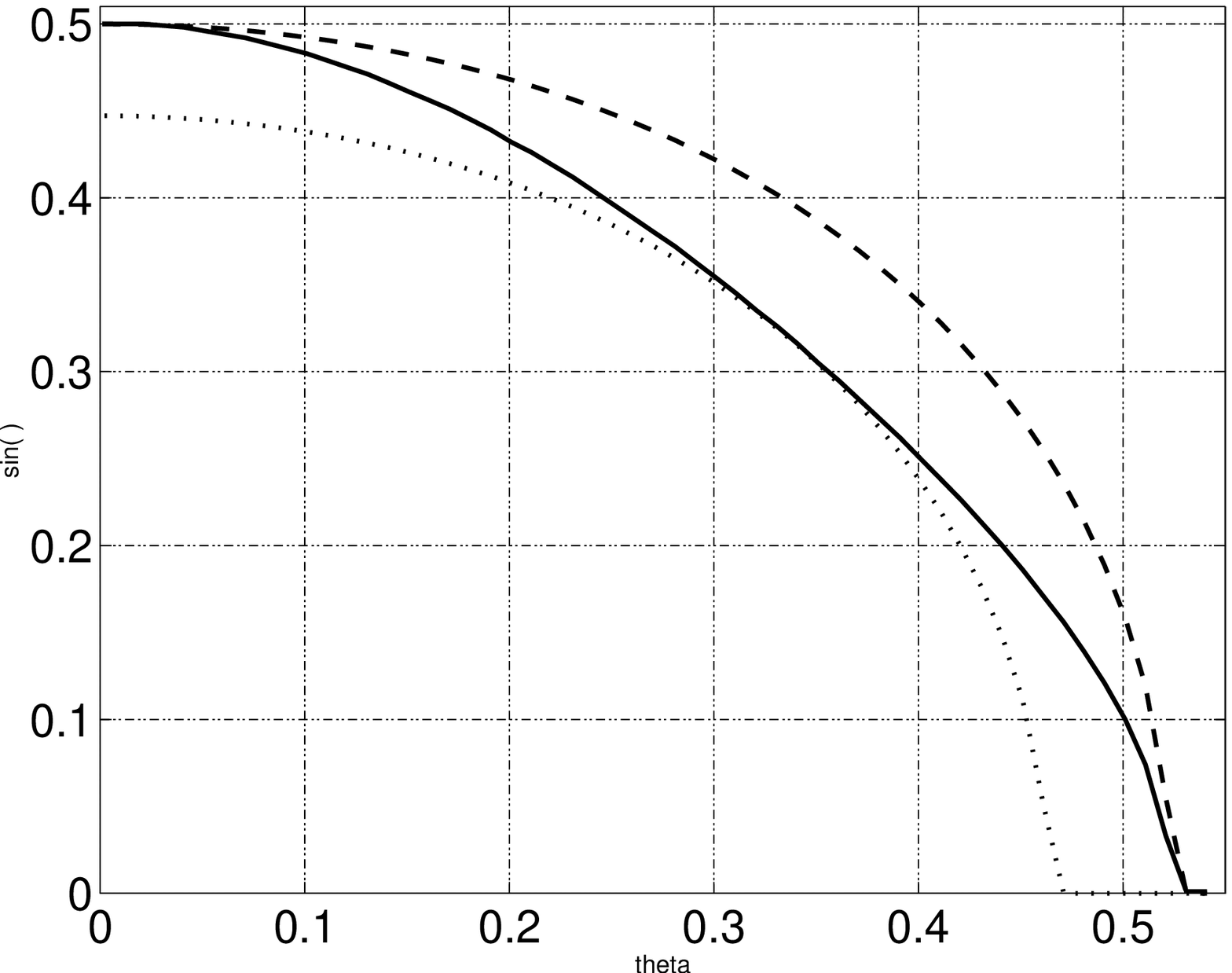}
\caption{A) Absolute value of the two eigenvalues $\nu_{1,2}$ of the 
rotated Jones matrix (solid red and black). Phase of the eigenvalues 
(divided by $\pi$) is plotted in dashes. The Eigenvalues become 
complex at the point where the two curves meet and join. This point 
lies between the Brewster angle (dashed vertical black) and the 
effective critical angle (dotted vertical black). Calculated for 
$\tan\theta=0.2$, $n=2$. B) (Black) The sine of the polarization 
critical angle PCA at which the eigenvalue of $R$ gets complex. 
(Dotted) Sine of the Brewster Angle, (Dashed) sine of the critical 
angle of total internal reflection. }
% The colorscale represents the amplitude of the resonance shift induced by the phase $\zeta$ of $\nu$.}
\label{fig:newCritical}
\end{figure}
Consider the discriminant of the eigenvalues, $D= 
\cos^2\!\xi\,(r_s-r_p)^2 + 4 r_s r_p$, which determines whether they 
are real or complex. Recall that the TM Fresnel reflection 
coefficient, $r_s$ is real and positive for all angles below the 
critical angle passing through unity and becoming complex and 
unimodular above the critical angle; whereas the TE reflection 
coefficient, $r_p$ is real below the critical angle but becomes 
negative at the Brewster angle and passes through negative one before 
becoming unimodular and complex above the critical angle.
It follows that $D$ will always be positive and $\nu_{1,2}$ real for 
angles of incidence below the Brewster angle. However for any 
non-zero value of the rotation angle $\xi$, $D$ will become zero {\it 
before} the critical angle since at the critical angle $D=4(1- 
\cos^2\!\xi)$ is negative. The value of the incidence angle $\xi$ 
when $D=0$ is the polarization critical angle (PCA) which we have 
already mentioned above.  Since it occurs when both $r_s,r_p$ have 
absolute value less than unity, the eigenvalues $\nu_{1,2}$ also have 
modulus less than unity and we have a phase shift upon reflection 
while a fraction $1- |\nu_{1,2}|^2$ of the incidence wave is 
refracted out.  This fraction can be calculated and does not 
correspond to either of the usual Fresnel transmission coefficients 
for TM or TE.  The behavior of the eigenvalues just described is 
shown in  Fig.~\ref{fig:newCritical}~A); the behavior of the 
polarization critical angle vs. $\sin \chi$ is shown in 
Fig.~\ref{fig:newCritical}~B).  We see that for small $\theta$ the 
onset of the the phase shift is close to the critical angle (CA); as 
$\theta$ varies the PCA moves close to the Brewster angle and then 
for $\theta$ close to the CA, where for any $\sin\chi$ we will have 
TIR, the PCA returns to the CA. This analysis allows us a simple 
understanding of why there is a PCA which precedes total internal 
reflection.  The TM and TE components of the spiral resonances have 
no relative phase shift at reflection until the Brewster angle; at 
the Brewster angle the TE component picks up a $\pi$ phase shift, 
which mixes with the TM component to give a phase shift between zero 
and $\pi$.  Right at the Brewster angle $r_p$ vanishes and the local 
TE component of the resonance is filtered out, but as the TE 
reflectivity picks up above the Brewster angle this phase shift 
appears before total internal reflection condition is reached. It is 
interesting to note that due to the finite phase aquired at the PCA, 
the resonance condition Eq.~\eqref{eq:planeapprox } is slightly 
changed. 
The term PCA suggests that at this angle the polarization 
properties of the spiral modes change.  We shall see that this is the 
case in the next section.

\section{Polarization properties of spiral modes}\label{sect:spiral}
\begin{figure}[t]
\centering
\psfrag{sinchi}{$\sin\chi$}
A)\includegraphics[width=7cm,height=4cm]{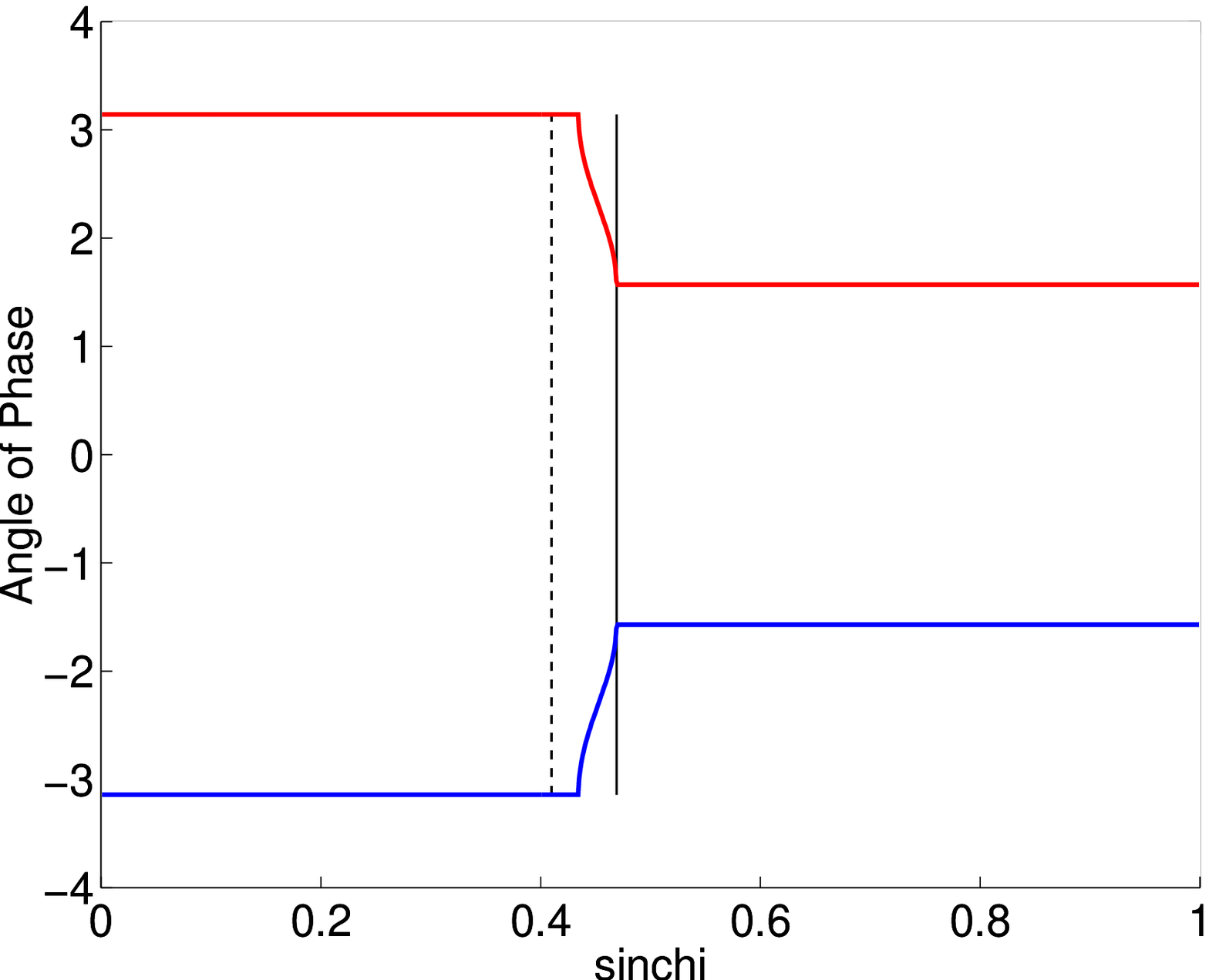}
\psfrag{Ratio}{$\left|\frac{nB_z}{E_z}\right|$}
B)\includegraphics[width=7cm,height=4cm]{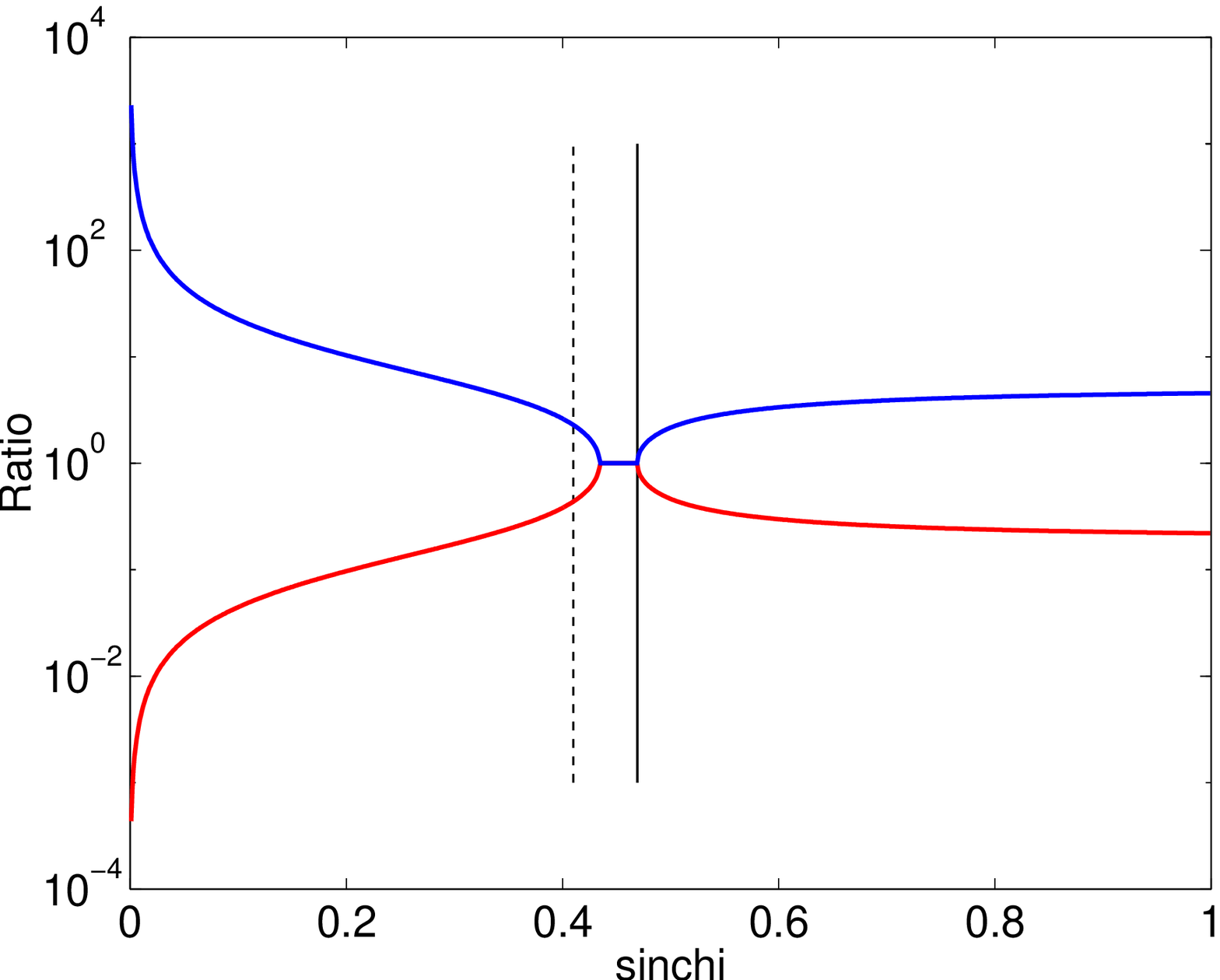}
\caption{A) Phase differences between the $E_z$ and $B_z$ field, for 
TE and TM like modes. B) Absolute values of the ratio of the 
eigenvector components of $R$. For normal incidence $\sin\chi=0$ the 
modes are clearly either TE or TM, right at the PCA they completely 
mix. Calculations done for $n=2$, $\tan\theta=0.2$. The solid 
vertical black line is the effective critical angle, the dashed line 
the effective Brewster angle.}
\label{fig:ebkPhaseRatio}
\end{figure}

When the eigenvalues of the $\mf{J}$ matrix are real, the 
eigenvectors can be chosen real (up to an overall phase) and thus 
there is no relative phase shift between $E_s$ and $E_p$.  While in 
general polarization is not well-defined inside the dielectric, in 
the short wavelength limit we are now examining it is well defined 
with respect to the considered ray direction, and zero relative phase 
shift implies linear polarization of the resonance with respect to 
the spiraling ray direction. Above the PCA, when the eigenvalues are 
complex, the eigenvectors also become complex and there is a relative 
phase shift between $E_s$ and $E_p$ corresponding to elliptical 
polarization of the internal field.  This picture is confirmed by the 
calculation of the eigenvalues plotted in 
Fig.~\ref{fig:ebkPhaseRatio} A), B).  At small $\sin \chi$ we have a 
large ratio between the eigenvector components and zero
phase shift, corresponding to linear polarization with the electric 
field almost completely in the $z$ or transverse direction (TM-like 
and TE-like).  After the PCA we get a phase shift approaching $\pi/2$ 
and equal ratios leading to circular polarization right at the CA; 
above the CA we have in general elliptical polarization for the 
internal fields with a calculable ellipticity and with a phase 
difference of $\pi/2$ as expected for whispering gallery modes.

The far-field polarization arising from this internal field can be 
found by applying to the relevant eigenvector of $\mf{J}$ the 
transmission matrix $J_t$ and then rotating the resulting vector by 
an angle $\Theta$ which projects the Jones vector onto the the plane 
perpendicular to the outgoing ray direction discussed in 
Section~\ref{sect:jones} above.  This angle $\Theta$ can be 
determined by straightforward geometric considerations which we omit 
here and simply state that the the far-field polarization vector 
$(E^{\prime}_s,E^{\prime}_p)$ is given by
\be
\mf{R}(-\Theta)J_t \ket{\mf{a}}
\ee
where $\mf{R}$ is the the rotation just mentioned and $\ket{\mf{a}}$ 
is either of the eigenpolarization vectors of $\mf{J}$.

\begin{figure}[t]
\centering
\psfrag{sinchi}{$\sin\chi$}
A)\includegraphics[width=7cm,height=4cm]{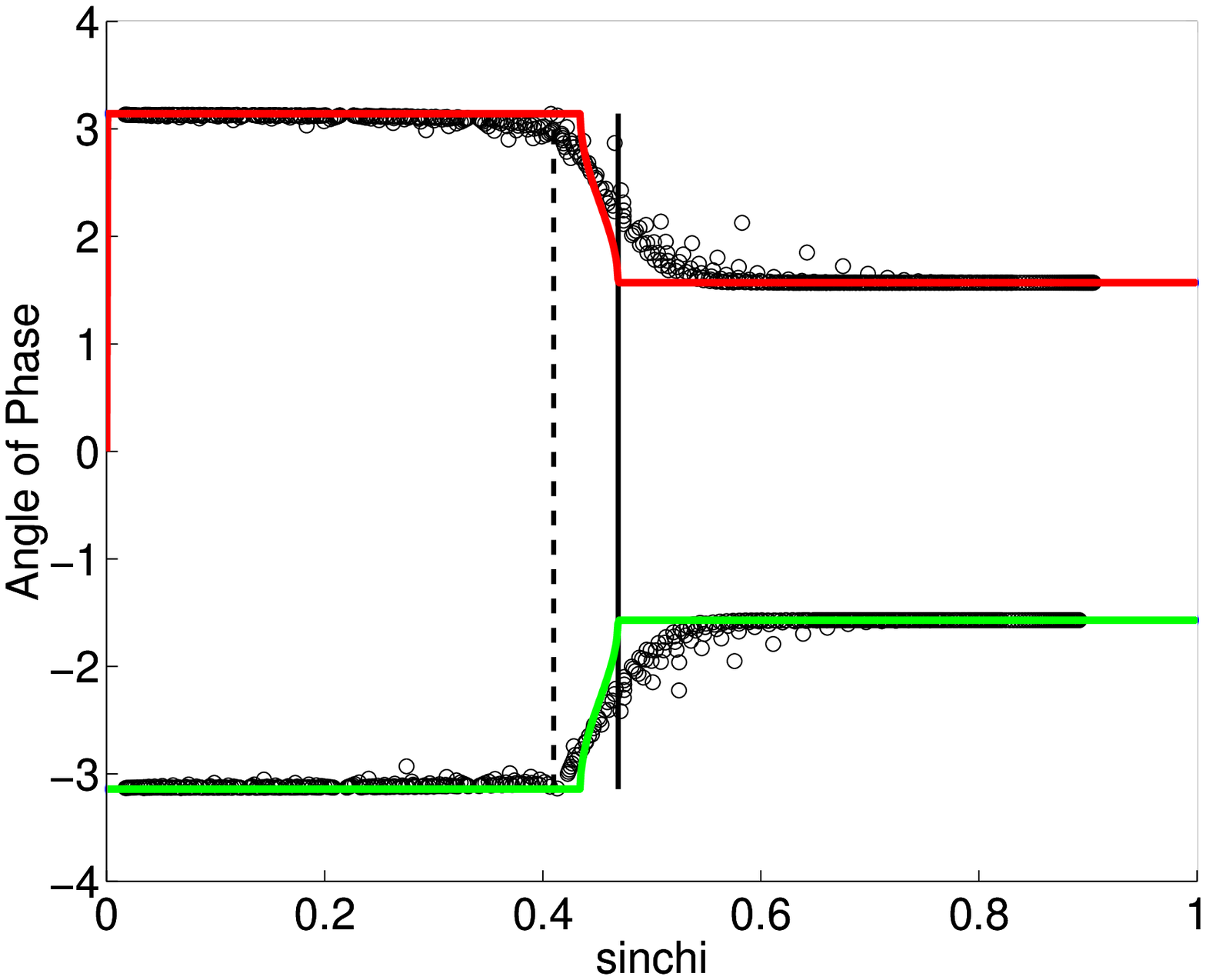}
\psfrag{Ratio}{$\left|\frac{nB_z}{E_z}\right|$}
B)\includegraphics[width=7cm,height=4cm]{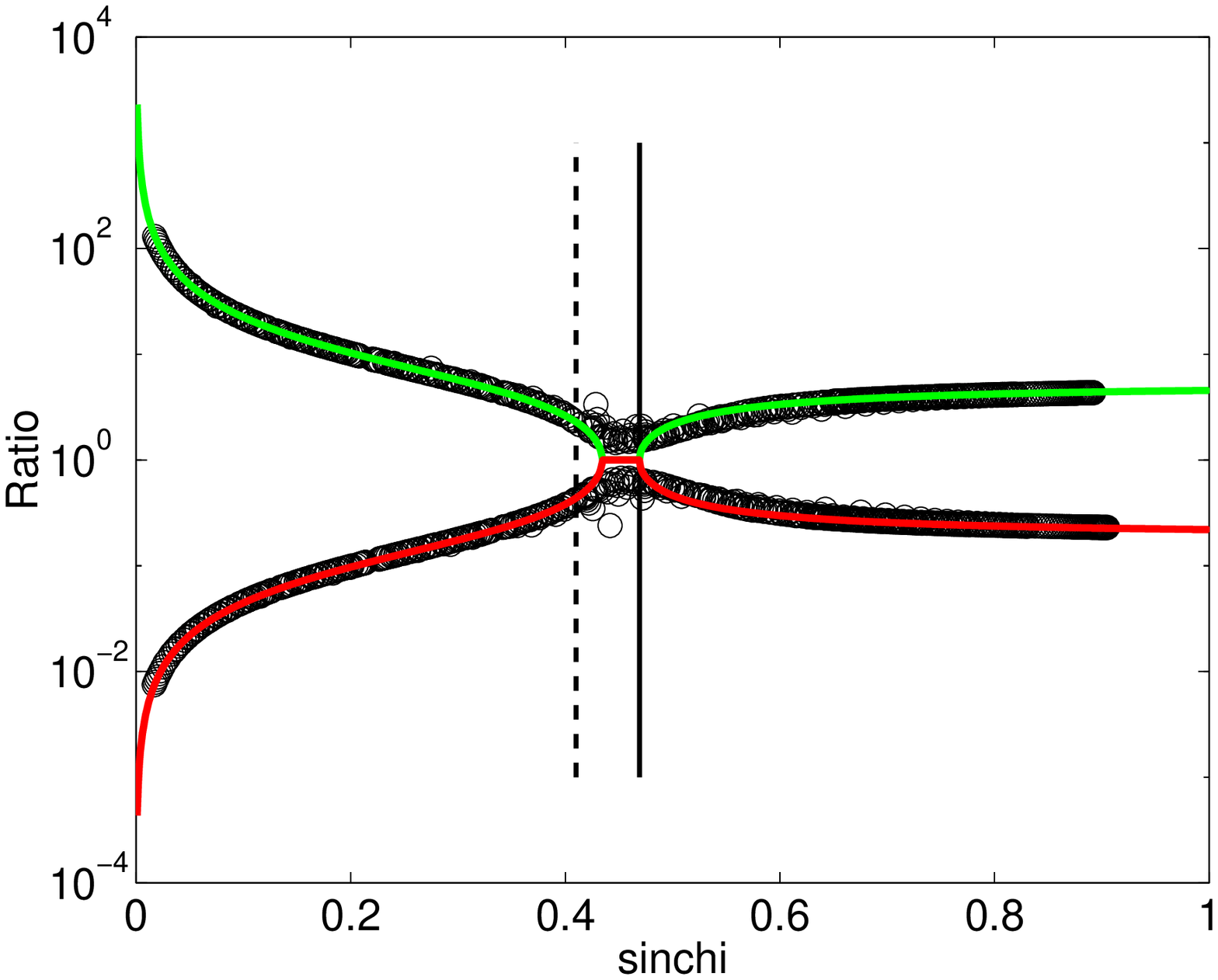}
\caption{Respective A) phase differences and B) absolute values of 
the ratio of the eigenvector components of $R$ (green and red solid) 
in the farfield. The black circles are the exact numerical solutions 
following Eq.~\protect\eqref{eq:3dresoCond} ($m\in[0,50]$ and 
$\gamma_1<50$). For $n=2$, $\tan\theta=0.2$. The solid vertical black 
line is the effective critical angle, the dashed line the effective 
Brewster angle.}
\label{fig:ebkPhaseRatioWave}
\end{figure}

In  Fig.~\ref{fig:ebkPhaseRatioWave} A), B), we compare the far-field 
polarization states for spiral resonances of the cylinder as 
predicted by the exact and geometric optics approach.  We find good 
agreement between the methods, although the exact solutions smooth 
the abrupt behavior near the PCA predicted by the geometric optics 
approach. Above the CA we have only evanecent emission, but the 
eccentricity of the two-vector is still finite. Note that as 
formulated, the Jones approach gives a continuous solution for the 
two polarization states, whereas the exact resonances are discrete 
and correspond to discrete allowed angles $\chi$ which can be also 
found through the EBK approach. The Jones approach provides a smooth 
and $k$-independent formula for the polarization states which agress 
with those of the resonaces.

\section{Summary and conclusions}\label{sect:summary }

We have reduced the Maxwell's equations for a dielectric rod of 
arbitrary cross-section to a vector Helmholtz equation for a 
two-component vector field living in the two-dimensional 
cross-sectional plane. We have devised a formulation of the resonance 
problem for the quasi-bound modes (spiral resonances), which can be 
implemented numerically for a general cross-section, and shown how 
the polarization state of the resonances in the farfield can be 
determined. Calculations were reported for the case of a circular 
cross-section (cylinder) and the results were compared to ray-optical 
results from an EBK formulation of the resonance problem in the 
semi-classical limit. We have analyzed the polarization state of the 
spiral resonances both inside the cylinder and in the farfield, 
and related its properties to the internal ray motion. It was shown 
that as the tilt angle of the spiraling ray with respect to the 
cross-sectional plane is increased, there exists a polarization 
critical angle at which the polarization changes from linear to 
elliptical both internally and externally and this occurs before the 
total internal reflection condition, so the effect can be measured 
readily in the far-field.  The physical picture we developed in terms 
of the Jones polarization vectors was useful in understanding the PCA 
and may be useful in generalizing the analysis to arbitrary 
cross-sections for which the ray motion can be chaotic. 
\section*{Acknowledgments}
We would like to thank Richard Chang and Andrew Poon for usefull discussions and greatfully acknowledge support from the National Science Foundation grant DMR-0408638. 

\bibliography{/home/harry/BIB/main.bib}    % BibTeX;  see file myrefs.bib

\begin{thebibliography}{10}

\bibitem{science98}
C.~Gmachl, F.~Capasso, E.~E. Narimanov, J.~U. N{\"o}ckel, A.~D. Stone,
  J.~Faist, D.~L. Sivco, and A.~Y. Cho.
\newblock High-power directional emission from microlasers with chaotic
  resonators.
\newblock {\em Science}, 280:1556--1564, 1998, cond-mat/9806183.

\bibitem{nature97}
J.~U. N{\"o}ckel and A.~D. Stone.
\newblock Ray and wave chaos in asymmetric resonant optical cavities.
\newblock {\em Nature}, 385:45--47, 1997, chao-dyn/9806017.

\bibitem{rex02}
N.~B. Rex, H.~E. Tureci, H.~G.~L. Schwefel, R.~K. Chang, and A.~D. Stone.
\newblock Fresnel filtering in lasing emission from scarred modes of
  wave-chaotic optical resonators.
\newblock {\em Phys.\ Rev. Lett.}, 88:art. no. 094102, 2002, physics/0105089.

\bibitem{Chern03}
G.~D. Chern, H.~E. Tureci, A.~D. Stone, R.~K. Chang, M.~Kneissl, and N.~M.
  Johnson.
\newblock Unidirectional lasing from {InGaN} multiple-quantum-well
  spiral-shaped micropillars.
\newblock {\em Appl.\ Phys.\ Lett.}, 83:1710--1712, 2003.

\bibitem{schwefel04}
Harald G.~L. Schwefel, Nathan~B. Rex, Hakan~E. Tureci, Richard~K. Chang,
  A.~Douglas Stone, Tahar Ben-Messaoud, and Joseph Zyss.
\newblock Dramatic shape sensitivity of directional emission patterns from
  similarly deformed cylindrical polymer lasers.
\newblock {\em J.\ Opt.\ Soc.\ Am.\ B}, 21:923--934, 2004, physics/0308001.

\bibitem{PoonCL98}
A.~W. Poon, R.~K. Chang, and J.~A. Lock.
\newblock Spiral morphology-dependent resonances in an optical fiber: {E}ffects
  of fiber tilt and focused {G}aussian beam illumination.
\newblock {\em Opt.\ Lett.}, 23:1105--1107, 1998.

\bibitem{Lock97b}
J.~A. Lock.
\newblock Morphology-dependent resonances of an infinitely long circular
  cylinder illuminated by a diagonally incident plane wave or a focused
  {G}aussian beam.
\newblock {\em J.\ Opt.\ Soc.\ Am.\ A}, 14:653--661, 1997.

\bibitem{RollS98}
G.~Roll and G.~Schweiger.
\newblock Resonance shift of obliquely illuminated dielectric cylinders:
  geometrical-optics estimates.
\newblock {\em Appl.\ Opt.}, 37:5628--5630, 1998.

\bibitem{schwefel05b}
Harald G.~L. Schwefel and A.~Douglas Stone.
\newblock Vector resonances in chaotic dielectric rods.
\newblock {\em in progress}, 2005.

\bibitem{tureci_thesis}
Hakan~E. T{\"u}reci.
\newblock {\em Wave chaos in dielectric resonators: {A}symptotic and numerical
  approaches}.
\newblock PhD thesis, Yale University, New Haven, USA, 2003.

\bibitem{schwefel_thesis}
Harald G.~L. Schwefel.
\newblock {\em Directionality and Vector Resonances of Regular and Chaotic
  Dielectric Microcavities}.
\newblock PhD thesis, Yale University, New Haven, USA, 2004.

\bibitem{HarayamaDI03}
T.~Harayama, P.~Davis, and K.~S. Ikeda.
\newblock Stable oscillations of a spatially chaotic wave function in a
  microstadium laser.
\newblock {\em Phys.\ Rev. Lett.}, 90:063901, 2003.

\bibitem{Tureci05}
H.~E. Tureci, H.~G.~L. Schwefel, Ph. Jacquod, and A.~Douglas Stone.
\newblock Modes of wave-chaotic dielectric resonators.
\newblock {\em Progress In Optics}, 47, 2005, physics/0308016.

\bibitem{slatec}
SLATEC.
\newblock {SLATEC} {Common} {Mathematical} {Library} ({Version} 4.1), July
  1993.
\newblock {\tt http://www.netlib.org/slatec/}.

\bibitem{Keller58}
J.~B. Keller.
\newblock Corrected {Bohr}-{Sommerfeld} quantum conditions for nonseparable
  systems.
\newblock {\em Ann. Phys.}, 4:180--188, 1958.

\bibitem{Keller60}
J.~B. Keller and S.~I. Rubinow.
\newblock Asymptotic solution of eigenvalue problems.
\newblock {\em Ann. Phys.}, 9:24--75, 1960.

\bibitem{einstein17}
A.~Einstein.
\newblock {Zum} {Quantensatz} von {Sommerfeld} und {Epstein}.
\newblock {\em Verhandl. Deut. Physik. Ges.}, 19:82--92, 1917.

\bibitem{jackson_book}
J.~D. Jackson.
\newblock {\em Classical electrodynamics}.
\newblock John Wiley \& Sons, Inc., New York, USA, 1998.

\bibitem{jones41}
R.~C. Jones.
\newblock A new calculus for the treatment of optical systems. {I.} description
  and discussion of the calculus.
\newblock {\em J.\ Opt.\ Soc.\ Am.}, 31:488--493, 1941.

\end{thebibliography}
\bibliographystyle{hunsrt}   % abbrvnat is in the natbib package

\end{document}